# Stiff-FCS: Single-Cell Stiffness Profiling With Integrated Molecular and Functional Analysis

Yuhao Zhang[a], Luyao Zhao[g], Zhengfu Huang[d], Kenan Song[d], Xianqiao Wang[d], Xianyan Chen[e], Jin Xie[g], Yiping Zhao[f], He Li[a], Leidong Mao[c], Yong Teng*[h, i], Yang Liu*[a, b]

[a]School of Chemical, Materials and Biomedical Engineering, College of Engineering, The University of Georgia, Athens, Georgia 30602, USA

[b]Institute of Bioinformatics, The University of Georgia, Athens, Georgia 30602, USA

[c]School of Electrical and Computer Engineering, College of Engineering, The University of Georgia, Athens, Georgia 30602, USA

[d]School of Environmental, Civil, Agricultural, and Mechanical Engineering, The University of Georgia, Athens, Georgia 30602, USA

[e]College of Public Health, The University of Georgia, Athens, Georgia 30602, USA

[f]Department of Physics and Astronomy, The University of Georgia, Athens, Georgia 30602, USA

[g]Department of Chemistry, The University of Georgia, Athens, Georgia, 30602, USA

[h]Department of Hematology and Medical Oncology, Winship Cancer Institute, Emory University, Atlanta, Georgia 30322, USA

[i]Wallace H. Coulter Department of Biomedical Engineering, Georgia Institute of Technology & Emory University, Atlanta, GA 30322, USA

*Email: Yang Liu (liuy@uga.edu); Yong Teng (yong.teng@emory.edu)

## ABSTRACT

Cell stiffness is a key determinant of how cells deform, migrate, and adapt to mechanically restrictive environments, yet existing single-cell stiffness assays remain difficult to combine with molecular analysis and downstream functional studies. To address these limitations, we introduce a microfluidic platform, stiffness-based ferrohydrodynamic cell sorting (Stiff-FCS), designed for high-throughput quantification of single-cell stiffness, on-chip molecular analysis, and post-assay cell recovery. Stiff-FCS combines ferrofluid-driven actuation with graded confinement channels to control cell movement, induce deformation, and spatially separate cells based on stiffness. An inverse computational model converts cell position and morphology into quantitative Young's modulus values. We demonstrate stiffness profiling of hundreds to thousands of cells per chip within minutes, same-cell fluorescence-based protein analysis, and recovery of stiffness-defined cells for downstream assays. Across diverse human and mouse cell lines, Lamin A/C showed the most consistent association with stiffness, whereas softer cells exhibited greater migratory capacity than stiffer cells. In a series of human head and neck cancer cell models, Stiff-FCS further resolved a stiff, less migratory subpopulation enriched in a higher-molecular-weight Vimentin state, offering a workflow for linking single-cell stiffness to molecular heterogeneity and cell behavior.

## INTRODUCTION

Single-cell mechanical phenotype is an important dimension of cell state as it determines how a cell deforms under mechanical force[1, 2, 3]. This property influences behaviors that depend on repeated shape change, including migration, invasion, circulation through confined spaces, and tissue remodeling[4, 5, 6]. Among the mechanical properties that govern cell behavior, stiffness is especially informative because it reflects a cell's resistance to deformation under external force[7]. Cell stiffness does not arise from a single molecular component, but from the combined contributions of the cytoskeleton, actomyosin tension, intermediate filaments, nuclear lamina, membrane properties, and intracellular organization[8, 9, 10]. In cancer, altered stiffness can influence how cells deform, detach, invade surrounding tissue, and migrate through confined environments, thereby contributing to invasive progression and metastatic dissemination[4, 11]. In immune cells, stiffness affects trafficking through the microvasculature, tissue infiltration, and movement through mechanically restrictive environments during immune surveillance or inflammation [12, 13, 14]. During development and tissue repair, cell stiffness also shapes coordinated cell movement and tissue remodeling, where repeated changes in cell shape are required for morphogenesis and regeneration[15, 16]. This heterogeneity is particularly important at the single-cell level, because bulk measurements can overlook rare but important cell states linked to migration, phenotypic plasticity, or therapeutic resistance[17, 18]. Single-cell stiffness profiling is therefore needed not only to measure deformability, but also to resolve functionally relevant cell states that are masked in population-averaged measurements.

Conventional methods such as atomic force microscopy (AFM), micropipette aspiration, and optical stretching have established the foundation of quantitative single-cell stiffness measurement. AFM can provide precise mechanical measurements with high spatial resolution, but it is generally low-throughput and labor-intensive.[19, 20, 21] Micropipette aspiration enables direct mechanical interrogation under controlled suction, yet it is technically demanding and difficult to scale to large and heterogeneous cell populations[22, 23]. Optical stretching offers contact-free assessment of deformability, but it also requires specialized instrumentation and is not readily coupled to downstream analysis of the same mechanically defined cells[24, 25]. More broadly, cells analyzed by these approaches are often difficult to recover for subsequent functional or molecular assays, creating a practical separation between mechanical measurement and biological follow-up. Thus, although conventional methods have been indispensable for defining cell mechanics, they are less well suited for scalable single-cell studies that aim to link stiffness with molecular state and downstream function.

Microfluidic technologies have substantially advanced single-cell mechanical phenotyping by enabling suspension-based analysis at higher throughput than conventional methods[26]. Current platforms can be broadly grouped into three categories. The first includes constriction- and filtration-based methods, represented by linearly tapered microchannels, gradual microfilters, and pore-based trapping devices, in which cells are forced through narrowing geometries and separated according to their resistance to deformation[27, 28, 29, 30]. These systems are conceptually simple and can generate strong mechanical contrast, but they often involve repeated cell-wall contact and remain susceptible to clogging, channel fouling, and incomplete recovery of mechanically defined cells. In addition, because cells deform as they pass through fixed physical gaps, the resulting readout is often influenced by both cell stiffness and size, making it difficult to isolate the intrinsic mechanical properties without size correction. The second category includes continuous hydrodynamic displacement methods, such as deformability-based deterministic lateral displacement and slanted-ridge or patterned-channel sorters, in which stiffness-dependent deformation alters cell trajectories under flow[31, 32, 33]. These platforms can achieve substantially higher throughput and continuous operation, but their performance depends strongly on device geometry, flow conditions, cell size, and mechanical relaxation, and they can still be affected by adhesion and fouling. The third category includes noncontact or low-contact methods, including extensional-flow stretching platforms and suspended microchannel resonators, which reduce hard surface contact and can preserve cell integrity during measurement[34, 35, 36]. However, these approaches are generally less suited for direct enrichment of large heterogeneous populations and are not readily integrated with downstream molecular analysis of the same mechanically defined cells. Detailed comparison of conventional microfluidic methods is provided in Supplementary Table S1. Collectively, these advances have moved the field toward more scalable single-cell mechanical phenotyping, but a practical gap remains between quantitative stiffness profiling, on-chip molecular characterization, and recovery of the same mechanically defined cells for follow-up studies.

To address these limitations, we developed a microfluidic platform that combines high-throughput single-cell stiffness quantification with on-chip molecular analysis and recovery of mechanically defined cells for downstream assays. A schematic overview of the full experimental workflow is provided in Supplementary Figure S1. The platform combines ferrofluid-driven actuation with graded confinement channels to control cell motion, induce deformation, and spatially separate cells according to stiffness. An inverse mechanical model was established to convert cell position and morphology into quantitative Young's modulus values. We term this approach stiffness-based ferrohydrodynamic cell sorting, or Stiff-FCS. Stiff-FCS can profile hundreds to thousands of cells per chip within approximately 3 min while remaining compatible with fluorescence-based protein analysis on the same cells. The workflow also preserves cell viability sufficiently for post-assay recovery and downstream functional and molecular studies, including migration assays and proteoform-resolved analysis. We validated Stiff-FCS across eight human and mouse cell lines and found that Lamin A/C showed the most consistent association with cell stiffness among the markers tested. We further showed that cells with higher migratory capacity exhibited lower stiffness compared to less migratory cells. Finally, we applied Stiff-FCS to patient-derived head and neck cancer models and, together with stiffness-resolved single-cell immunoblotting, identified enrichment of a higher-molecular-weight Vimentin state, VIM′, in stiffer subpopulations. These results establish Stiff-FCS as a scalable framework for linking single-cell stiffness to molecular heterogeneity and downstream cell behavior.

## RESULTS & DISCUSSION

### Overview of the Stiff-FCS approach

The goal of the Stiff-FCS method is to provide quantitative information about the single-cell stiffness in terms of Young's modulus with a high throughput and high precision. In this approach, a mixture of target cells and colloidal, stable magnetic fluid (ferrofluid) is loaded into a polydimethylsiloxane (PDMS) device patterned with multiplexed, parallel single-cell elastic-grading channels. Cells will first be aligned at the entrance of the grading channels. A magnetic field gradient is then applied along the channel axis, inducing cell movement via magnetic buoyancy arising from the magnetization difference between the cell and the surrounding ferrofluid (**Figure 1a, b**).

The cell is treated as a deformable, initially spherical body with a diameter $D$, immersed in a viscous medium under low Reynolds number conditions (Re < 0.001). Inertia and channel deformation are neglected. Cell deformation occurs due to geometric confinement when the available channel across the section is smaller than the undeformed cell diameter. The friction or non-specific binding between the cell and PDMS channel walls is assumed to be negligible due to the lubrication with 0.5% (w/v) bovine serum albumin and 2 mM EDTA. Under these conditions, the dominant forces acting on the cell along the channel axis are the magnetic buoyancy force, the hydrodynamic drag force during motion, and the elastic compression force due to cell deformation.

**1) Magnetic buoyancy force in ferrofluid.** The magnetic buoyancy force is generated from the ferrofluid under a non-uniform magnetic field, due to the ferrofluid's magnetic nanoparticle-induced pressure imbalance on the cell's surface, and directs the cells towards the magnetic field minima. The magnetic force on diamagnetic particles (e.g., tumor cells) in ferrofluid under an external magnetic field has been reported in previous studies[37, 38, 39, 40, 41, 42]. The expression of the magnetic buoyancy force is

$$\vec{F}_m = -\mu_0 V\{(\vec{M}_{ferrofluid} - \vec{M}_{cell}) \cdot \nabla\}\vec{H} \tag{1}$$

where $\mu_0$ is the permeability of free space, $V$ is the cell volume, $\vec{M}_{ferrofluid}$ is the magnetization of the ferrofluid surrounding the cell, $\vec{M}_{cell}$ is the magnetization of the cell, and $\vec{H}$ is the magnetic field strength at the center of the cell. For a cell suspended in ferrofluids under a strong magnetic field, the magnetization of the ferrofluid containing superparamagnetic particles depends on the ferrofluid concentration and can be modeled by the Langevin function[37]. Because the magnetization of biological cells is negligible compared with that of the ferrofluid, the force is dominated by the ferrofluid term and acts to drive the cell toward regions of lower magnetic field strength. In the Stiff-FCS device, the magnetic buoyancy force provides a continuous axial driving force that is independent of cell deformation state. The force magnitude can be adjusted by varying the ferrofluid concentration and magnetic field configuration.

**2) Hydrodynamic drag during cell motion.** During motion through the channel, the cell experiences a viscous drag force from the surrounding ferrofluid. Under low Reynolds number conditions, the drag force can be written as:

$$\vec{F}_d = -3\pi\eta D\left(\vec{U}_{cell} - \vec{U}_{ferrofluid}\right) f_D \quad (2)$$

where the drag force is proportional to the cell diameter, ferrofluid viscosity ($\eta$), and velocity difference between cell ($\vec{U}_{cell}$) and ferrofluid ($\vec{U}_{ferrofluid}$) . $f_D$ is the hydrodynamic drag force coefficient for a cell moving near the solid surface[43, 44, 45].

**3) Elastic compression force from geometric confinement.** As cells enter the narrow grading channel with a size smaller than their diameter, geometric confinement induces cell deformation through contact with the channel walls. The resulting elastic compression is governed by linear elasticity or large deformation mechanics, depending on the degree of deformation. In this framework, the cell is treated as a deformable body compressed symmetrically by rigid walls. Cell stiffness can be estimated using the Hertz model, the Tatara model, or the hyperelastic Tatara model, depending on the deformation regime[46]. The elastic compression force $\vec{F}_{el}$ is related to cell deformation through a linear relationship that accounts for both non-spherical geometry after deformation and intrinsic hyperelasticity of cells. Here, we define the deformation ratio as x:

$$x = 1 - \frac{D_{deform}}{D} \quad (3)$$

where $D_{deform}$ is the apparent deformed cell diameter measured along the direction perpendicular to compression, as illustrated in the Supplementary Figure S2.

For small deformation (x < 0.1), the cell's Young's modulus $E_{cell}$ can be estimated using the Hertz model[47 48]:

$$E_{cell} = \frac{1\left(1 - v^2\right)\vec{F}_{el}}{\sqrt{D}\left(D - D_{deform}\right)^{3/2}} \tag{4}$$

where $v$ is the Poisson's ratio of the cell. $v = 0.5$ is used in this model.

For moderate deformation (0.1 < x <0.3), deformation-induced geometric flattening must be considered, and the $E_{cell}$ can be estimated with the Tatara model:[49]

$$E_{cell} = \frac{3\left(1 - v^2\right)\vec{F}_{el}}{2\left(D - D_{deform}\right)a} - \frac{2\vec{F}_{el}}{\pi\left(D - D_{deform}\right)f(a)} \tag{5}$$

where $a$ is the contact radius and $f(a)$ is a characteristic length that accounts for the non-spherical cell geometry after deformation. These terms are defined as:

$$a = \frac{1}{2}\left(\sqrt{D^2 - D_{deform}^2} + D_{deform} - D\right) \tag{6}$$

$$f(a) = \left[\frac{(1 + v)D^2}{2\left(a^2 + D^2\right)^{3/2}} + \frac{1 - v^2}{\left(a^2 + D^2\right)^{1/2}}\right]^{-1} \tag{7}$$

Compared with the Hertz model, the Tatara model incorporates deformation-induced geometric flattening and therefore provides a more accurate description of cell compression in the moderate deformation regime. A Poisson ratio of $v = 0.5$ was used in this model.

As cells undergo large deformation (x > 0.3), neither the Hertz model nor the conventional Tatara model is sufficient to fully describe the compression behavior, because both pronounced geometric flattening and intrinsic material nonlinearity must be considered. Under this condition, the cell can be treated as a hyperelastic body symmetrically compressed by rigid walls, and the cell Young's modulus $E_{cell}$ can be estimated using the hyperelastic Tatara model[50, 51]:

$$E_{cell} = \frac{3A\left(1-\nu^2\right)\vec{F}_{el}}{2a\left(D-D_{deform}\right)}\left(1+2B\frac{a^2}{D_{deform}^2}\right) - \frac{2A\vec{F}_{el}}{\pi(D-D_{deform})f(a)}\left(1+4B\frac{a^2}{5D_{deform}^2}\right) \quad (8)$$

where A and B are hyperelastic correction factors, given by

$$A = \frac{(1-x)^2}{1-x+x^2/3},\ B = \frac{1-x/3}{1-x+x^2/3},\ x = 1-\frac{D_{deform}}{D} \quad (9)$$

Here, x denotes the deformation ratio, while *a* and *f(a)* retain the same definitions as those used in the Tatara mode. By incorporating both geometry and constitutive nonlinearities, the hyperelastic Tatara model provides a more appropriate description of cell deformation under strong confinement. A Poisson ratio of $\nu = 0.5$ was used in this model.

**4) Forces at balanced position.** As the cell moves through the grading channel, the magnetic buoyancy force drives it forward, while elastic compression from geometric confinement resists its motion. Once the cell reaches its equilibrium position, its velocity becomes zero, and the viscous drag force vanishes. At this point, the cell no longer advances because the driving magnetic force is balanced by the restoring elastic compression force generated by deformation against the channel walls. Since the cell is compressed symmetrically by two channel walls, the elastic compression force acts along the two contact directions. The component of the magnetic buoyancy force along each wall normal is therefore balanced by the corresponding elastic restoring force. As a result, the force balance at equilibrium can be written as

$$\vec{F}_{el} = \frac{\vec{F}_m}{2\sin(\theta)} \quad (10)$$

This force balance determines the final stopping position of the cell inside the grading channel. By combining this relation with the Hertz, Tatara, or hyperelastic Tatara model, depending on the deformation regime, the cell stiffness can be estimated from the observed cell deformation at equilibrium.

**Design Principle of Stiff-FCS**

The theoretical framework described above establishes several key requirements for the design of the Stiff-FCS platform. First, the magnetic buoyancy force must exceed the elastic compression resistance at the entrance of the grading channel so that cells can be driven into the constricted region. Second, cells with heterogeneous sizes, including those smaller than the local channel width, must still be stably positioned within the device for stiffness profiling. Third, a large number of parallel grading channels is required to enable high-throughput single-cell analysis. Fourth, cells distributed across different equilibrium positions should remain accessible for downstream collection and analysis after the stiffness assay.

Based on these considerations, we designed the prototype Stiff-FCS device shown in **Figure 1c, d**. The device integrates a cell loading channel, a cell collection channel, and a screening region composed of 265 parallel grading channels. A permanent magnet is positioned beneath the device using a custom 3D printed holder that enables precise control of the distance between the magnet and the screening region (Supplementary Figure S3). The magnet length exceeds the total width of the screening region, which helps maintain a relatively uniform magnetic driving force across parallel channels and minimizes channel-to-channel variation in cell actuation. Each grading channel has a constant height of 10 μm, while the channel width decreases monotonically from $W_{in}$ = 31 μm to $W_{out}$ = 7 μm over a channel length of 1150 μm, corresponding to a tapering angle of 1.2 º. This geometry establishes a continuous confinement gradient, enabling cells with different mechanical properties to migrate to distinct equilibrium positions based on their deformability. To ensure robust handling of cells with heterogeneous diameters, we introduced circular resting sites along each grading channel (Supplementary Figure S4). These structures serve two purposes. First, they improve retention of smaller cells during sample loading by providing local trapping regions prior to magnetic actuation. Second, they facilitate subsequent cell transfer and retrieval for downstream analysis. In addition, the collection channel is positioned parallel to the loading channel to balance hydrostatic pressure and enable controlled initiation and cessation of ferrofluid flow through the screening region. The collection channel also provides an accessible outlet for recovery of the softest cells, which migrate furthest under magnetic actuation.

This device architecture enables parallel screening of 100-1000 cells per assay while preserving positional information related to single-cell stiffness. With the current channel geometry and magnetic configuration, the Stiff-FCS platform supports stiffness measurements over a broad range, from approximately 0.01 to 100 kPa. Importantly, the design also permits downstream recovery of selected cells for functional assays and molecular profiling, enabling direct linkage between cell mechanics and biological state.

To generate sufficient magnetic actuation for cell deformation within confined channels, we optimized the magnetic field configuration of the Stiff-FCS platform (**Figure 2**). A 100 μm thick glass substrate was used to minimize the distance between the permanent magnet and the screening channels, thereby maximizing the magnetic field and its spatial gradient at the microchannel array (**Figure 2a**). Under this configuration, simulations showed a magnetic flux density of up to 0.9 T and a magnetic flux density gradient as high as 3146 T $m^{-1}$ near the bottom edge of the cell loading channel (**Figure 2b-d**). The magnetic flux density and corresponding gradient were highest near the loading side and decreased progressively toward the opposite end of the device (**Figure 2e–g**). This field distribution establishes a strong magnetic buoyancy force on diamagnetic cells suspended in ferrofluid, directing them away from the high-field region and toward the low-field region, namely the cell collection channel.

## Calibration of Stiff-FCS for Stiffness-Based Cell Positioning and Single-Cell Stiffness Calculation

To calibrate Stiff-FCS for quantitative stiffness readout, we first defined how magnetic actuation governs the steady-state cell position in the device. In Stiff-FCS, cells are spatially separated according to their mechanical response under magnetic buoyancy forcing, while single-cell stiffness is inferred from the final equilibrium position in the grading channel together with the corresponding cell morphology at that position (**Figure 3a**). To make the platform adaptable to different cell types, we examined two key operating parameters that control magnetic actuation, namely the relative distance between the magnet and the cell loading channel along the y direction, and the ferrofluid concentration. Human non-small cell lung cancer H1299 cells were used for calibration. These cells spanned an apparent diameter range of 5 to 45 μm[40]. Cells smaller than 7 μm were excluded from analysis because they were smaller than the narrowest channel width ($W_{out}$ = 7 μm), and therefore could pass through the full screening channel without being retained. Using a three-dimensional (3D) numerical model implemented in MATLAB, we simulated the final cell position, denoted as $Y$, under different operating conditions. Increasing the magnet-channel distance reduced the magnetic buoyancy force and shifted the final cell position toward lower Y values, indicating a shorter moving distance. This trend was confirmed experimentally (**Figure**

**3b**). We therefore selected a magnet distance of 0 μm for subsequent studies to maximize magnetic actuation and dynamic range. Ferrofluid concentration provided a second means to tune the magnetic buoyancy force by modulating ferrofluid magnetization according to Eq. 1. Simulations predicted that increasing ferrofluid concentration increased the final cell position and promoted greater cell migration distance. Experimental measurements showed the same overall dependence (**Figure 3c**). On the basis of these results, we selected 0.3% (v/v) ferrofluid as the working condition, which provided strong magnetic actuation while limiting potential adverse effects of higher ferrofluid concentrations on cell viability[52]. Cell viability studies confirmed that 97.11% of cells remained viable after this assay (Supplementary Figure S5) using 0.3% (v/v) ferrofluid. In addition, post-separation growth analysis suggested that the collected cells retained proliferative capacity after processing through Stiff-FCS (Supplementary Figure S6). Together, these results established the magnetic operating parameters required for robust, biocompatible Stiff-FCS measurements.

We next studied how intrinsic cell properties (e.g., stiffness and diameter) shape equilibrium positions under fixed operating conditions: a magnetic distance of 0 μm and a ferrofluid concentration of 0.3% (v/v). Simulation results confirmed that cell final positions are strongly dependent on cell stiffness. For a representative cell (diameter: 15 μm), increasing stiffness will shift the final position toward smaller Y values. This is consistent with the greater resistance of stiffer cells to channel confinement and the increased elastic compression force (**Figure 3d**). Chemical fixation is known to induce covalent crosslinking of intracellular proteins, thereby increasing cellular stiffness[53]. To validate the mechanical sensitivity of Stiff-FCS, we analyzed paraformaldehyde (PFA)-fixed cancer cell lines. Fixed cells exhibited significantly reduced migration distances within the device compared to fresh controls, consistent with increased stiffness under confinement-based sorting (Supplementary Figure S7). The cell final position was also strongly influenced by cell diameter. For a fixed stiffness of 1kPa, increasing the cell diameter reduced the moving distance, because larger cells experienced stronger geometric confinement within the grading channel (**Figure 3e**). These results demonstrate that the cell final position is jointly determined by cell mechanics and cell geometry. We therefore mapped the final position as a function of both stiffness and cell diameter under the optimized operation conditions (**Figure 3f**). This two-parameter landscape defined the operating regime of the Stiff-FCS device and identified the region in which stiffness could be inferred when cell morphology was considered in the analysis.

Based on the calibrated positional map, we next established an inverse mechanical workflow for rapid stiffness calculation. For each cell on the Stiff-FCS, the measured final position (X, Y) and morphology were used as direct experimental inputs. These measurements were then integrated with the local magnetic field, ferrofluid concentration, local channel dimensions, and cell deformation metrics, including apparent deformed cell diameter and aspect ratio, in a stiffness calculator built on the steady state force balance framework described above. Depending on the degree of deformation, the stiffness calculator applies the Hertz, Tatara, or hyperelastic Tatara model to solve the problem inversely and compute the corresponding Young's modulus value of each cell. The importance of this stiffness calculator is that it transforms Stiff-FCS from a positional sorting assay into a quantitative mechanics analysis platform. Automating the conversion of cell position and morphology into stiffness enables rapid, standardized, and scalable analysis of large cell populations. This substantially simplifies stiffness profiling and makes high-throughput single-cell mechanical phenotyping more practical for routine biological studies.

## Validation of Stiff-FCS for Multiparametric Single-Cell Mechanical Phenotyping Across Diverse Cell Types

Having established the calibration framework and inverse stiffness calculator, we next validated Stiff-FCS as a multiparametric single-cell phenotyping platform using diverse cancer and stromal cell lines. We first used human breast cancer cell line MDA-MB-231 as a representative model. We asked whether Stiff-FCS could capture single-cell mechanical heterogeneity alongside fluorescence-based protein expression. Immunofluorescence imaging performed directly in the device confirmed simultaneous visualization of Vimentin, Lamin A/C, and DAPI after stiffness-based sorting (**Figure 4a**). Application of the stiffness calculator yielded quantitative single-cell stiffness values for each cell (**Figure 4b**). Analysis of 1029 individual MDA-MB-231 cells revealed a broad stiffness distribution across the population, demonstrating that Stiff-FCS can resolve substantial mechanical heterogeneity within a single cell line. We further benchmarked Stiff-FCS against AFM using cells recovered from the top collection channel and the bottom loading channel (Supplementary Figure S8). AFM measurements confirmed that cells retrieved from the collection channel showed lower Young's modulus values than cells recovered from the loading channel, consistent with the soft and stiff fractions defined by Stiff-FCS. This agreement in relative mechanical ranking supports the validity of the Stiff-FCS stiffness readout. It is worth noting that the absolute Young's modulus values obtained by Stiff-FCS were lower than those measured by AFM. This difference is likely attributable to the distinct mechanical states and loading geometries probed by the two methods. AFM was performed on adherent cells attached to a rigid glass substrate, a condition that can increase apparent stiffness through bottom effect, cell spreading, and adhesion-mediated contractility[54, 55, 56]. By contrast, Stiff-FCS measures cells in suspension under geometric confinement, which more closely reflects the mechanics of rounded non-adherent cells. In addition, AFM and Stiff-FCS probe different deformation modes, boundary conditions, and spatial scales, and therefore their absolute modulus values are not expected to match directly. Vimentin is a major intermediate filament protein involved in cytoskeletal organization and cell shape remodeling, whereas Lamin A/C is a key component of the nuclear lamina that governs nuclear architecture and mechanics[57, 58, 59]. Consistent with these structural roles, single-cell stiffness showed weak positive correlations with equivalent cell diameter and Vimentin intensity (Spearman $\rho = 0.15$ and 0.17, respectively; both $P < 0.001$) and a modest positive correlation with Lamin A/C intensity (Spearman $\rho = 0.27$, $P < 0.001$) (**Figure 4c–e**). These results are consistent with partial coupling of cell stiffness to cell size, cytoskeletal organization, and nuclear architecture, while also indicating that stiffness cannot be fully explained by any single feature alone.

We extended the validation of the Stiff-FCS assay using a total of 8 cell lines: human breast cancer cell lines (MCF7, HCC1806, HCC70), human non-small cell lung cancer (NSCLC) cell lines (H1299, H3122), mouse triple-negative breast cancer cell line (PY8119), and human stromal cell line (Fibroblast). Distinct distributions of final position and corresponding single-cell stiffness were resolved across these cell lines (**Figure 4f, g**). Fibroblasts exhibited the highest stiffness (median: 22.15 kPa, n = 568 cells), while PY8119 and H3122 cells were among the softest (median: 0.17 kPa, n = 1297 cells; and median: 0.25kPa, n = 1336 cells, respectively), with the other cell lines occupying intermediate mechanical states. This trend was mirrored by final position distributions and further supports the accuracy of the Stiff-FCS force-balance framework. Each cell line also displayed substantial within-line spread, highlighting the intrapopulation mechanical heterogeneity. Additionally, Stiff-FCS provided morphology-derived descriptors for the same cells, including aspect ratio and solidity (**Figure 4h, i**). Although these features varied across cell lines, they did not provide a uniform explanation for the observed stiffness differences across populations. For example, fibroblasts showed the largest aspect ratio, consistent with a more elongated morphology, but this did not correspond to increased deformability. This result indicates that cell shape anisotropy and mechanical stiffness are related but not equivalent properties. Solidity varied only modestly across cell lines, further suggesting that simple shape descriptors alone are insufficient to account for the mechanical differences captured by Stiff-FCS.

To examine whether expression of structural proteins was linked to cell stiffness, we investigated Vimentin and Lamin A/C expression in the same cells. Across the tested cell panel, Lamin A/C showed a more consistent positive association with stiffness than Vimentin, particularly among the human cancer cell lines (**Figure 4j, k** and Supplementary Table S2). Positive stiffness-Lamin A/C correlations were observed in MDA-MB-231 (Spearman $\rho = 0.27$, $P < 0.001$), MCF7 (Spearman $\rho = 0.34$, $P < 0.001$), HCC1806 (Spearman $\rho = 0.38$, $P < 0.001$), HCC70 (Spearman $\rho = 0.21$, $P < 0.01$), H1299 (Spearman $\rho = 0.14$, $P < 0.01$), and H3122 cells (Spearman $\rho = 0.28$, $P < 0.001$), whereas the relationship between stiffness and Vimentin was weaker and more variable across cell types. In particular, Vimentin showed only weak positive correlations in some lines (MDA-MB-231 (Spearman $\rho = 0.17$, $P < 0.001$), HCC70 (Spearman $\rho = 0.14$, $P < 0.05$), H1299 (Spearman $\rho = 0.19$, $P < 0.001$), Fibroblast (Spearman $\rho = 0.22$, $P < 0.001$)), no clear association in others (MCF7 ($P > 0.05$), HCC1806 (Spearman $\rho = 0.12$, $P > 0.05$), PY8119 (Spearman $\rho = 0.02$, $P > 0.05$)), and an inverse relationship in H3122 cells (Spearman $\rho = -0.26$, $P < 0.001$). These results indicate that Lamin A/C is a more consistent correlate of the stiffness measured by Stiff-FCS than Vimentin, although neither marker alone fully accounts for the observed mechanical heterogeneity. This interpretation was further supported by analysis of cells recovered from the loading and collection channels (**Figure 4l, m**). In multiple cell lines, the stiffer loading-channel fraction tended to exhibit higher Lamin A/C intensity than the softer collection-channel fraction, whereas Vimentin differences between fractions were less consistent. These findings support a closer coupling between nuclear structural organization and the stiffness state measured by Stiff-FCS, while also indicating that cellular stiffness remains an integrated phenotype shaped by multiple structural determinants.

We finally examined whether the integrated single-cell measurements generated by Stiff-FCS could support higher-dimensional phenotypic mapping. t-SNE and principal component analysis (PCA) were performed using cell final position (Y), diameter, stiffness, aspect ratio, solidity, Vimentin intensity, and Lamin A/C intensity for each cell line (**Figure 4n, o**). In both analyses, cells within each cell line were distributed into structured patterns rather than diffuse clouds, suggesting that the Stiff-FCS measurements capture coordinated variation across multiple single-cell features. When cells were colored by log-transformed stiffness, cells with different stiffness values occupied different regions of the maps, indicating that stiffness varies together with other measured single-cell features. The geometry of these distributions differed across cell lines, consistent with cell-type-specific relationships among mechanics, morphology, and structural protein expression. In PCA, PC1 explained 37.6–54.5% of the variance across the tested cell lines, whereas PC2 explained 18.6–35.8% (**Figure 4o**). PCA loading analysis further showed that the relative contributions of Vimentin, Lamin A/C, final position, diameter, and stiffness to PC1 and PC2 varied across cell lines (Supplementary Figure S9), arguing against a universal single-feature explanation of the observed heterogeneity. These results show that Stiff-FCS generates information-rich single-cell data that support not only quantitative stiffness profiling, but also multidimensional phenotypic analysis.

### Application of Stiff-FCS Reveals an Association Between Reduced Stiffness and Enhanced Cell Migration

Cell mechanical properties are closely linked to migratory behavior, as the reduced stiffness can facilitate cell deformation and shape adaptation during movement[60, 61]. However, direct comparison of the stiffness of migratory and non-migratory subpopulations at the single-cell level remains challenging using conventional approaches. Our lab has designed a two-dimensional migration assay, termed Race-Track, that monitors dynamic cell migration at the single-cell level and enables recovery of subpopulations based on migration behavior[62]. After 24 h of migration toward 10% fetal bovine serum (FBS) as a chemoattractant, we collected the subpopulations with the longest migration distances (migratory) and the shortest migration distances (non-migratory) from Race-Track for subsequent Stiff-FCS analysis (**Figure 5a**). Across fibroblasts and multiple cancer cell lines, migratory cells consistently exhibited lower stiffness than their non-migratory counterparts (**Figure 5b**). The shift was significant ($P < 0.001$) across the tested cell lines and was most evident in Fibroblast, H1299, HCC1806, and MDA-MB-231. The median stiffness of migratory cells was reduced by approximately 1.36- to 7.65-fold relative to the matched non-migratory populations across these lines. MCF7 cells displayed minimal migratory behavior in Race-Track, and therefore, the migratory subpopulation could not be robustly collected for downstream stiffness analysis. Notably, these migratory-prone cell lines also showed elevated Vimentin expression in the validation dataset (**Figure 4j**), consistent with the established association of Vimentin with migratory phenotypes[62]. Given the weaker and more variable coupling between Vimentin and stiffness across cell lines, this pattern suggests that Vimentin may reflect migratory programs more directly than the stiffness state measured by Stiff-FCS.

To test this relationship in the reverse direction, we first separated cells according to stiffness using Stiff-FCS and then evaluated their migratory behavior in Race-Track (Supplementary Figure S10). In this reciprocal experiment, soft subpopulations migrated faster than stiff subpopulations, further supporting a functional association between reduced stiffness and increased migration capacity. These results indicate that mechanical compliance is a defining feature of migratory cell states in this system.

**Mechanical State-Resolved Migratory and Vimentin Proteoform Heterogeneity in Patient-Derived Head and Neck Cancer (HNC) Cells**

HNC exhibits pronounced heterogeneity in epithelial-mesenchymal plasticity, migratory behavior, and immune escape[63, 64, 65]. Because these processes are accompanied by changes in cytoskeletal organization, nuclear mechanics, and immune-regulatory signaling, we asked whether stiffness-resolved phenotyping could uncover biologically meaningful cell states in patient-derived HNC models. We therefore applied Stiff-FCS to four human HNC cell lines, JHU029, SCC47, HN6, and HN12. Stiff-FCS resolved distinct single-cell stiffness distributions across four HNC models (**Figure 6a, b**, Supplementary Table S3, providing detailed values) and revealed substantial inter-line mechanical heterogeneity. The median stiffness values were 0.40, 1.23, 0.59, and 1.04 kPa for JHU029, SCC47, HN6, and HN12, respectively. Among these specimens, HN12 displayed the broadest and right-shifted stiffness distribution, whereas JHU029 and HN6 were comparatively softer, with SCC47 occupying an intermediate range. In parallel, we measured the expression of Vimentin, Lamin A/C, epidermal growth factor receptor (EGFR), and programmed death-ligand 1 (PD-L1) across four HNC specimens (**Figure 6c-f**). Vimentin intensity was higher in JHU029 and HN12, Lamin A/C was elevated in SCC47 and HN12, EGFR showed strong expression in SCC47, and PD-L1 was most abundant in JHU029. These results indicate that the mechanically distinct HNC states are associated with greater molecular heterogeneity.

We next examined whether cell migration in patient-derived HNC models was associated with altered mechanical state. Race-Track measurements revealed clear differences in migration behavior across the four HNC lines, with JHU029 showing the greatest average migration distance (515.92 ± 340.24 μm) and SCC47 the lowest (240.22 ± 146.91 μm), whereas HN12 (252.22 ± 229.96 μm) and HN6 (332.19 ± 238.84 μm) displayed intermediate behavior (**Figure 6g**). To determine whether these behavioral differences were accompanied by mechanical differences at the subpopulation level, we isolated migratory and non-migratory cells from each line and profiled them using Stiff-FCS (**Figure 6h**). Across all four HNC models, migratory cells consistently exhibited lower stiffness than the corresponding non-migratory cells (**Figure 6i**). In JHU029, the mean stiffness of migratory cells was 0.39 ± 0.32 kPa, compared with 1.62 ± 1.51 kPa in non-migratory cells. The corresponding values were 0.78 ± 1.24 versus 4.24 ± 10.25 kPa for SCC47, 0.60 ± 0.63 versus 2.31 ± 2.66 kPa for HN6, and 0.73 ± 0.77 versus 2.97 ± 3.72 kPa for HN12. This difference was significant in every line ($P < 0.001$). These results indicate that enhanced migration in patient-derived HNC cells is linked to a softer mechanical state and suggest that reduced stiffness was consistently observed in migratory HNC subpopulations across the tested lines.

We next asked whether stiffness-resolved molecular profiling could reveal features that were not apparent from total fluorescence intensity alone. To address this, we isolated stiff and soft HNC subpopulations by Stiff-FCS and analyzed them using single-cell western blotting for Vimentin and PD-L1 (**Figure 6j, k**)[66]. This assay resolved two Vimentin isoforms, operationally defined here as VIM′ and VIM″, while simultaneously mapping PD-L1 intensity at single-cell resolution through the point color scale (**Figure 6l**). Across all four patient-derived HNC lines, stiff cells showed reproducible enrichment of VIM′-positive states relative to matched soft cells. When the VIM′+/VIM″+ and VIM′+/VIM″− quadrants were combined, the total VIM′-positive fraction increased from 2.3% to 23.2% in JHU029, from 9.2% to 60.3% in SCC47, from 5.4% to 42.7% in HN6, and from 4.1% to 57.1% in HN12 when comparing soft and stiff fractions, respectively. Conversely, the VIM′−/VIM″+ population was reduced from 80.4% to 59.8% in JHU029, from 90.1% to 39.6% in SCC47, from 80.0% to 43.5% in HN6, and from 76.7% to 40.0% in HN12. These data indicate that stiffness is associated with a marked redistribution of Vimentin proteoform states, rather than with total Vimentin abundance alone. This proteoform shift was biologically coherent with the migration results. In the same HNC models, migratory cells were softer than non-migratory cells, whereas stiff subpopulations were consistently enriched for VIM′. In JHU029, SCC47, and HN12, this shift was dominated by expansion of the VIM′+/VIM″+ compartment, whereas in HN6, there was also a substantial increase in VIM′+/VIM″− cells. This pattern is consistent with VIM′ enrichment in a mechanically stiffer HNC cell state that also shows reduced migratory capacity. It also explains why total Vimentin fluorescence alone was insufficient to fully resolve stiffness states in the preceding analyses. PD-L1 intensity also varied across the Vimentin isoform-defined subpopulations, indicating that the stiffness-resolved HNC states were potentially accompanied by additional immune-relevant molecular heterogeneity.

In addition to the overall increase in VIM′-positive cells, stiffness-resolved HNC subpopulations also differed in the relative abundance of combined Vimentin/PD-L1 molecular states (**Figure 6m, n**). Across all four lines, the soft fractions were enriched in the VIM′−/VIM″+ / PD-L1+ state, accounting for 67.2%, 7.2%, 35.5%, and 75.3% of JHU029, SCC47, HN6, and HN12 soft cells, respectively. By contrast, stiff fractions showed redistribution toward one or more VIM′+ states, although the dominant VIM′+ configuration differed by line. In JHU029 and HN12, stiff cells were enriched predominantly in the VIM′+/VIM″+/PD-L1+ compartment, which increased from 2.3% to 17.4% and from 3.9% to 34.2%, respectively, when comparing soft and stiff fractions. In SCC47, the stiff fraction was enriched mainly in the VIM′+/VIM″+ PD-L1− state, whereas in HN6 the stiff fraction expanded both the VIM′+/VIM″−/PD-L1− and VIM′+/VIM″+/PD-L1− compartments. Z-score analysis summarized this shift across lines and showed depletion of the VIM′−/VIM″+ / PD-L1+ state in stiff cells together with enrichment of one or more VIM′+ states (**Figure 6n**). These data indicate that stiffness-resolved HNC subpopulations differ not only in total Vimentin proteoform abundance, but also in the composition of combined Vimentin / PD-L1 molecular states. PD-L1 intensity itself also differed between stiff and soft fractions, but the direction of change was not conserved across the four lines (**Figure 6o**). JHU029 showed a modest increase in PD-L1 intensity in the stiff fraction, SCC47 showed only a limited shift, whereas HN6 and HN12 showed higher PD-L1 intensity in the soft fraction. Thus, although PD-L1 contributes to the molecular heterogeneity of stiffness-resolved HNC subpopulations, absolute PD-L1 intensity alone does not provide a consistent readout of mechanical state in this exploratory panel. Instead, the most reproducible cross-line feature of the stiff fraction was enrichment of VIM′-associated states, indicating that Vimentin proteoform composition tracks stiffness more robustly than total PD-L1 abundance.

This work establishes Stiff-FCS as a quantitative and scalable platform for single-cell stiffness profiling across diverse biological contexts. Beyond quantitative stiffness profiling, Stiff-FCS can be coupled with morphology, migration assays, fluorescence imaging, and stiffness-resolved single-cell western blotting to resolve mechanically distinct cell states. In patient-derived HNC cells, this integrated workflow identified a stiff, less migratory, VIM′-enriched subpopulation, demonstrating that stiffness-resolved proteoform analysis can reveal molecular heterogeneity that is not captured by bulk marker intensity alone. Several directions may further extend the utility of Stiff-FCS, including improving throughput and dynamic range, expanding multiplexed molecular readouts, adapting the workflow for rare or limited clinical specimens, and integrating complementary downstream assays with stiffness-resolved cells. Because cells can be recovered after profiling, Stiff-FCS is also well positioned for future studies linking mechanical state to cell function over time, therapeutic response, and multi-omic cellular identity.

## METHODS

**Ethical statement.** This research complies with all relevant ethical regulations. Human head and neck cancer cell lines used in this study were obtained from established research sources and collaborators as de-identified laboratory materials. No identifiable human participant information was accessed by the authors. Accordingly, these materials were handled in accordance with institutional policies for research use of de-identified human-derived biological materials.

**Statistics and reproducibility.** No statistical method was used to predetermine sample size. Single-cell measurements in Stiff-FCS are reported as individual cell-level observations unless otherwise stated, and independent experiments were performed using separate devices and independently prepared samples, as indicated in the figure legends. Statistical analyses were performed on independent experimental replicates where applicable. Correlation analyses used Spearman's rank correlation. Box plots indicate the median, interquartile range, and whiskers extending to 1.5 times the interquartile range. Investigators were not blinded during sample handling or device operation. However, cell position, morphology, stiffness, and most image-derived features were quantified using predefined or automated analysis pipelines. Data were excluded only in cases of clear technical failure based on predefined quality-control criteria.

**Model of Stiff-FCS and its validation.** We developed an analytical model used in this study to simulate 3D cell trajectories in ferrofluids.[67, 68] It could predict 3D transport of diamagnetic cancer cells and magnetic WBCs in ferrofluids inside a microfluidic channel coupled with arbitrary configurations of permanent magnets. The magnets produced a spatially non-uniform magnetic field that led to a magnetic force on the cells. Trajectories of the cells in the device were obtained by (1) calculating the 3D magnetic force via an experimentally verified and analytical distribution of magnetic fields as well as their gradients, together with a nonlinear Langevin magnetization model of the ferrofluid and (2) solving governing equations of motion using analytical expression of magnetic force and hydrodynamic viscous drag force in MATLAB (MathWorks Inc., Natick, MA).

**Microfluidic Device Fabrication.** The ferrohydrodynamic–microfluidic platform was designed using AutoCAD (Autodesk, San Rafael, CA, USA) and fabricated via multilayer SU-8 soft lithography. SU-8 3005 (Kayaku Advanced Materials Inc., Westborough, MA, USA; NC2744251) and SU-8 3050 (Kayaku Advanced Materials Inc., Westborough, MA, USA; NC0702369) were sequentially patterned using a two-step photolithographic process to generate structures with distinct heights. The moving channels were defined with a height of 10 μm and a width linearly tapered from 31 μm to 7 μm, whereas the loading channels were fabricated with a height of 50 μm and a width of 200 μm. The fabricated moving channel height was measured to be 10.2 μm using a stylus surface profilometer (Dektak 150, Bruker, Tucson, AZ, USA), confirming consistency with the designed thickness.

To facilitate device release, the silicon master was rendered superhydrophobic by vapor-phase silanization with trichloro(1H,1H,2H,2H-perfluorooctyl) silane (97%, Sigma-Aldrich, 448931) under vacuum for 30 min. Polydimethylsiloxane (PDMS; Sylgard 184 Silicone Elastomer Kit, Dow) was prepared at a 10:1 (base:curing agent) ratio, degassed under vacuum, cast onto the treated master, and cured at 80 °C for 2 h to obtain the elastomeric replica. The cured PDMS slab was peeled from the master and bonded to microscope cover glasses (24 × 50 mm, No. 1, Fisherbrand, 12541056) following air plasma activation in a low-pressure RF plasma cleaner (Harrick Plasma, PDC series) operated at high power for 45 s. The bonded assemblies were subsequently heated at 80 °C for 15 min to promote interfacial siloxane condensation and strengthen the bond, yielding the assembled ferrohydrodynamic–microfluidic platform. The bonded microfluidic chip was then mounted onto a custom-designed 3D-printed holder incorporating one NdFeB permanent magnet (N52, K&J Magnetics, Pipersville, PA,

USA), forming the integrated ferrohydrodynamic–microfluidic platform

**Ferrofluid synthesis and characterization.** A water-based γ-iron oxide ($\gamma$-$Fe_2O_3$) nanoparticle ferrofluid was synthesized using a standard chemical co-precipitation method. For biocompatibility, the suspension pH was adjusted to 7.0 using sodium hydroxide (NaOH, Sigma–Aldrich, S5881), and osmolarity was balanced by supplementation with Hank's Balanced Salt Solution (HBSS, Thermo Fisher Scientific, 14025092). Nanoparticle morphology and size distribution were characterized by transmission electron microscopy (TEM; FEI, Eindhoven, The Netherlands), yielding a mean particle diameter of 10.91 ± 4.86 nm. Magnetic properties were measured using a vibrating sample magnetometer (VSM; MicroSense, Lowell, MA, USA). Magnetization curves were fitted to the Langevin function, resulting in a saturation magnetization of 1,107 A $m^{-1}$ and a magnetic volume fraction of 0.3% (v/v). Working ferrofluid concentrations of 0.2% and 0.1% (v/v) were prepared by dilution with phosphate buffered saline (Gibco PBS, pH 7.4, Thermo Fisher Scientific, 10010023).

**Cell culture.** Human NSCLC cell lines, H1299 and H3122, human breast cancer cell lines, MDA-MB-231, MCF-7, HCC70 and HCC1806, murine breast cancer cell line PY8119, and human primary fibroblasts were obtained from the American Type Culture Collection (ATCC, Manassas, VA, USA). Human HNC cell lines, HN6 and HN12 cells, were a kind gift from Dr. Andrew Yeudall (Augusta University, Augusta, GA); JHU029 cells were obtained from Dr. David Sidranski at Johns Hopkins University (Baltimore, MD), and UM-SCC-47 (SCC47) cells were kindly provided by Dr. Georgia Chen at Emory University, (Atlanta, GA). All cell lines were used for experiments before passage 10 and cultured in the Dulbecco's Modified Eagle Medium (DMEM, high glucose, Thermo Fisher Scientific, 11965092) or RPMI-1640 medium (Thermo Fisher Scientific, 11875093) supplemented with 10% (v/v) fetal bovine serum (FBS, Thermo Fisher Scientific, A5670701), 0.1 mM non-essential amino acids solution (NEAA, Thermo Fisher

Scientific, 11140050), and 1% (v/v) penicillin–streptomycin solution (Thermo Fisher Scientific, 15140122). Cells were dissociated using 0.05% trypsin–EDTA (Thermo Fisher Scientific, 25300054) or 0.25% trypsin–EDTA (Thermo Fisher Scientific, 25200072) at 37 °C for 5 min. Cells were centrifuged at 300g for 3 minutes, and an appropriate amount of PBS or ferrofluid was added to achieve a concentration of approximately one million cells per milliliter, cell concentration was determined by using an automated cell counter (Countess, Thermo Fisher Scientific).

**Preparation of PFA-fixed cells for loading.** MDA-MB-231 and HCC1806 cells were cultured under standard conditions and harvested at a concentration of approximately $1 \times 10^6$ cells/mL. The cell suspension was centrifuged at 300g for 3 min, and the supernatant was removed. The cell pellet was then fixed with 4% PFA at room temperature for 10 min with gentle mixing. After fixation, cells were centrifuged again and washed twice with 1× phosphate-buffered saline (PBS) to remove residual fixative.

**Cell Stiffness experiment setup and procedure.** Prior to experiments, microfluidic devices were passivated with 0.5% (w/v) bovine serum albumin (BSA; Sigma-Aldrich, A9647) and 2 mM EDTA (Invitrogen, 0.5 M, pH 8.0; 15575020) in 1× PBS. Devices were centrifuged to remove trapped air and incubated overnight at 4 °C to minimize nonspecific cell adhesion. After removal of excess solution, ferrofluid at the desired concentration was introduced and centrifuged again to eliminate residual bubbles. Subsequently, 3 μL of ferrofluid or PBS containing cells at a concentration of approximately $1 \times 10^6$ cells $mL^{-1}$ was loaded into each inlet. A slight hydrostatic height difference was maintained between inlet and outlet reservoirs to prevent cell trapping within the loading channels. Devices were mounted on the stage of an inverted microscope (Axio Observer, Carl Zeiss) for real-time imaging. Bright-field images and videos were acquired using a CCD camera (Carl Zeiss). After initial cell positioning, the device was placed onto a 3D-printed holder incorporating NdFeB permanent magnets (N52, K&J Magnetics). The magnet position was adjusted as required, and cell behavior under the applied magnetic field was recorded.

**Single-cell migration assay.** Cells were maintained under standard culture conditions (37 °C, 5% $CO_2$). After centrifugation, cells were resuspended in serum-free medium at approximately $1 \times 10^6$ cells $mL^{-1}$. To promote cell adhesion, the microfluidic channels were coated with 50 µg $mL^{-1}$ collagen type I (Corning, 354236) overnight at 4 °C. Before cell loading, excess collagen solution was removed, and the channels were rinsed with PBS and equilibrated with serum-free medium at 37 °C for 30 min. Cells were then introduced into the cell loading channel. By applying a transient hydrostatic height difference, cells were guided toward the entrances of the migration channels, where they aligned in a single-file arrangement along the channel axis. After cell seeding, the device was incubated at 37 °C and 5% $CO_2$ for 40 min to allow cell attachment to the glass surface. A bright-field image was acquired at the beginning of the assay (t = 0 h) to confirm that cells were positioned at the entrances of the migration channels as the common starting locations for migration analysis.

A chemotactic gradient was established by introducing medium containing FBS into the chemokine channel while maintaining serum-free medium in the loading channel. The device was then kept in the incubator (37 °C, 5% $CO_2$) for the duration of the migration assay. Media in the corresponding reservoirs were refreshed every 6–8 h to maintain gradient stability. At the end of the 24 h assay, the device was imaged again in bright-field mode to record the final positions of cells. Cells that remained in the loading channel or failed to enter the migration channels were defined as non-migratory. Cells that entered and traversed the migration channels toward the chemokine side were defined as migratory. The two populations were then collected separately from the device and transferred into a 96-well plate for downstream cell stiffness analysis.

For migration analysis, the initial cell positions were defined based on the t = 0 h bright-field image, in which cells were confirmed to be located at the entrances of the migration channels. Final cell positions were determined from the endpoint images. Migration distance was defined as the straight-line displacement from the channel entrance to the final position of each cell along the migration axis, and migration speed was calculated by dividing the migration distance by the total assay time (24 h). This analysis assumes that all cells started from the channel entrance and migrated predominantly along the channel direction.

**On-chip Immunofluorescence staining procedure.**

Unlike the previously described surface coating strategy used to minimize cell–wall interactions, microfluidic channels were coated with collagen type I (50 ug/mL) overnight at 4 °C to promote cell adhesion. Excess collagen solution was removed, and channels were rinsed with PBS and allowed to dry prior to ferrofluid loading. Ferrofluid at a concentration of 0.3% (v/v) was introduced into the microchannels and degassed to remove residual bubbles. Cells were then loaded and allowed to adhere to the collagen-coated surfaces by incubation at 37 °C and 5% $CO_2$ for 10-15 min.

Immunostaining was performed entirely within the microchannels. Cells were fixed with 4% paraformaldehyde (Thermo Fisher Scientific, J61899.AK) for 15 min at room temperature and washed three times with PBS (5 min each). Samples were permeabilized with 0.1% Triton X-100 (Sigma-Aldrich, T8787) for 15 min and washed again with PBS. Non-specific binding was blocked using UltraCruz® Blocking Reagent (Santa Cruz Biotechnology, sc-516214) for 30 min at room temperature.

Primary antibodies, including anti-Vimentin (Abcam, ab92547), anti-Lamin A/C (Abcam, ab238303), anti-PD-L1 (Abcam, ab205921) and anti-EGFR (Abcam, ab52894), were introduced and incubated for 1 h at room temperature. After PBS washes, fluorescently labeled secondary antibodies donkey anti-rabbit Alexa Fluor™ Plus (Invitrogen, A32790) and donkey anti-mouse Alexa Fluor™ Plus (Invitrogen, A32787) were applied for 1 h at room temperature. Following additional PBS washes, nuclei were stained with the DAPI Fluoromount-G® reagent (SouthernBiotech, 0100-20) according to the manufacturer's instructions, and devices were imaged by fluorescence microscopy.

**Live/Dead Assay.** Microfluidic channels were coated with collagen type I overnight at 4 °C. Ferrofluid (0.3%, v/v) was introduced and degassed to remove trapped air. Cells were loaded into the device and allowed to adhere to the collagen-coated channel surfaces by incubation at 37 °C and 5% $CO_2$ for 10-15 min prior to viability staining. Cell viability was evaluated on-chip using Calcein AM (Invitrogen, C3099) and ethidium homodimer-1 (EthD-1; Invitrogen, E1169). A staining solution was prepared in DPBS (Gibco, 14190144) to final concentrations of 2 μM Calcein AM and 4 μM EthD-1. The solution was introduced into the microchannels to ensure complete filling without trapped air and incubated for 30 min at 37 °C under light-protected conditions. Following incubation, devices were directly imaged by fluorescence microscopy. Live cells exhibited intracellular green fluorescence, whereas dead cells displayed red nuclear fluorescence.

**Proliferation Assay.** Following the Cell Stiffness experiment setup and procedure, cells were recovered separately from the loading channel and the collection channel by flushing each region with complete culture medium using independent pipette tips to avoid cross-contamination. Cells collected from the loading channel were designated as stiff cells, whereas cells recovered from the collection channel were designated as soft cells. To minimize carryover of ferrofluid, each cell suspension was transferred to a microcentrifuge tube and centrifuged; the supernatant containing residual ferrofluid was carefully aspirated without disturbing the pellet, and the cells were resuspended in fresh complete medium. The wash step was repeated as needed until the supernatant appeared optically clear. Equal volumes of the resulting stiff and soft cell suspensions were then seeded into 96-well tissue culture plates and maintained under standard culture conditions (37 °C, 5% $CO_2$). Cell growth and morphology were monitored longitudinally, and proliferation was assessed by recording each condition at 1, 3, 5, and 7 days after seeding.

**Cell Stiffness measurement using AFM.** Cell mechanical properties were quantified by atomic force microscopy (NX10, Park Systems, Korea) under liquid conditions using an established protocol[69]. Measurements were performed in 1× PBS using QP-Bio AC cantilevers (nominal tip radius <10 nm; spring constant ~0.06 N $m^{-1}$ in liquid). Before measurement, cells collected from Stiff-FCS were seeded onto poly-L-lysine-coated (poly-L-lysine hydrobromide, Sigma-Aldrich, P2636) glass substrates and allowed to attach in a humidified incubator for 15 min. Deflection sensitivity was calibrated on a rigid glass surface, and the cantilever spring constant was determined in liquid using the thermal noise method. For each condition, approximately 10 cells were analyzed, and three force–distance curves were acquired per cell from the perinuclear cytoplasmic region while avoiding the cell edge and nuclear center. Curves were collected at an approach velocity of ~2 μm $s^{-1}$ to minimize rate-dependent viscoelastic effects. The maximum loading force was limited to ~1 nN, resulting in a typical indentation depth of ~300–400 nm and thereby reducing substrate influence. Young's modulus was obtained by fitting the loading portion of the force–indentation curves to the Hertz model, assuming a Poisson's ratio of 0.5. Curves lacking a clear contact point, showing abnormal baseline drift or adhesion artefacts, or yielding poor Hertz fits were excluded. The remaining curves were averaged to generate one modulus value per cell. All measurements were performed on live cells.

**Single-cell Western Blot.** Polyacrylamide (PA) gel slides containing microwell arrays were fabricated using an SU-8 patterned silicon mold. An 8%T precursor solution was prepared from 30% acrylamide/bis-acrylamide (29:1; Sigma-Aldrich, A3699), benzophenone methacrylamide (BP-APMA; 3-Benzoyl-N-[3-(2-methyl-acryloylamino)-propyl]benzamide, CAS 1706951-11-2; Raybow USA, Brevard, NC, USA), 1× Tris–glycine buffer (Sigma-Aldrich, T4904), and deionized water.

Following degassing, ammonium persulfate (0.08% w/v; Sigma-Aldrich, A3678) and TEMED (0.08% v/v; Sigma-Aldrich, T9281) were added to initiate polymerization. The precursor was cast between the SU-8 mold and a silanized glass slide and polymerized for ~20 min at room temperature. Gels were released and stored in deionized water until use. For single-cell loading, a suspension (~$1 \times 10^6$ cells $mL^{-1}$ in PBS) was applied to the microwell array and allowed to settle for ~10 min. Excess cells were removed by PBS rinsing to ensure predominantly single-cell occupancy. Cell lysis and electrophoresis were performed in a heated lysis/electrophoresis buffer (55 °C) containing SDS, sodium deoxycholate, Triton X-100 (Sigma-Aldrich, T8787), and Tris–glycine buffer. Cells were lysed for 30 s prior to electrophoretic separation. Proteins were separated under an electric field of 240 V $cm^{-1}$ for 15 s and immediately immobilized by UV-mediated photocapture (350-360 nm, ~1.8 J $cm^{-2}$, 45 s). Gels were washed extensively in TBST.

In-gel immunoprobing was performed using primary antibodies diluted 1:100 in 2% (w/v) BSA (Sigma-Aldrich, A2153). Rabbit monoclonal anti-vimentin and mouse monoclonal anti–PD-L1 (1073628-5, Abcam, AB210931) were incubated for 2 h at room temperature. After washing, fluorophore-conjugated secondary antibodies (1:100 dilution; donkey anti-rabbit Alexa Fluor™ Plus and donkey anti-mouse Alexa Fluor™ Plus) were applied for 1 h under light-protected conditions. Gels were washed, desalted in deionized water, and dried under nitrogen prior to imaging. Protein expression was quantified by extracting electrophoretic band intensity profiles at the single-cell level. Antibody concentrations were optimized to ensure linear signal response within the dynamic detection range.

**Acknowledgements:** This work was funded by the National Institutes of Health (NIH) 1R01GM16324. The study was also supported in part by Georgia CTSA/Regenerative Engineering and Medicine (REM) Pilot Grants Award (Y.L and Y.T.).

**Author contributions:** Y.L. designed the microfluidic platform (Stiff-FCS). Y.L., Y.T., and YH.Z. designed the experiments. YH.Z. performed stiffness assay on the system optimization, validation, and application. Y.L., XQ.W., and H.L. designed software and performed data analysis. YP.Z. provided guidance on hybrid system setup. Y.T. provided guidance on cancer-related applications and edited the manuscript. XY.C. provided statistical analysis support. ZF.H. performed AFM measurements. LY.Z. performed single-cell proteoform separation and analysis. KN.S. supported the design and fabrication of the magnet-device holder. LD.M. guided ferrofluid synthesis and magnetophoresis in the channel. Y.L. wrote the manuscript.

**Competing interests:** The authors declare no competing interests.

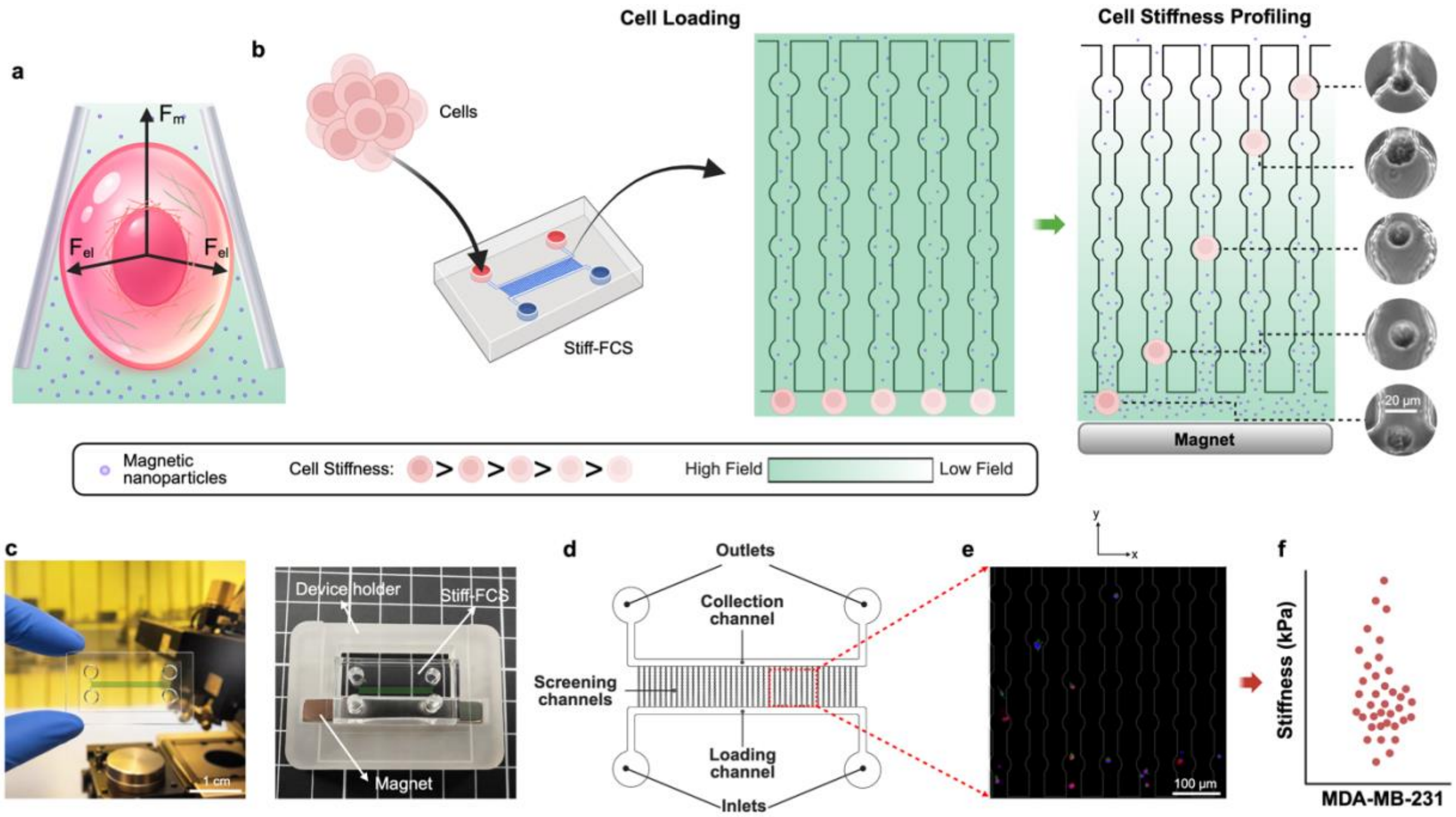

**Figure 1. Overview of the stiffness-based ferrohydrodynamic cell sorting platform (Stiff-FCS) and prototype device.** (a) Schematic illustration of the forces acting on a cell in a screening micro-track of the Stiff-FCS device. Cells are subjected to a magnetic buoyancy force ($F_m$) generated by the ferrofluid under an applied magnetic field and an elastic compression force ($F_{el}$) arising from geometric confinement within the micro-track. (b) Conceptual workflow of stiffness measurement using Stiff-FCS. Cells are introduced into the loading channel and aligned at similar initial lateral positions before entering parallel screening micro-tracks. The micro-tracks are designed with widths smaller than the cell diameter to impose confinement. Under an applied magnetic field, cells are driven into the screening micro-tracks and migrate until the magnetic buoyancy force and elastic compression force are balanced. Cells with lower stiffness experience reduced elastic resistance and travel farther into the device, whereas stiffer cells remain closer to the loading side. Representative phase-contrast images of cells during migration are shown at right. (c) Photographs of the prototype Stiff-FCS chip and experimental setup. A custom holder is used to control the relative position between the microfluidic chip and the permanent magnet. (d) Device layout (top view) of the Stiff-FCS platform, consisting of a loading channel, a collection channel, and parallel screening micro-tracks. (e-f) Representative fluorescence image of cells in the screening micro-tracks and the corresponding single-cell stiffness distribution obtained from final cell position and morphology. The platform enables stiffness profiling at single-cell resolution and can be integrated with downstream molecular measurements on the same cells.

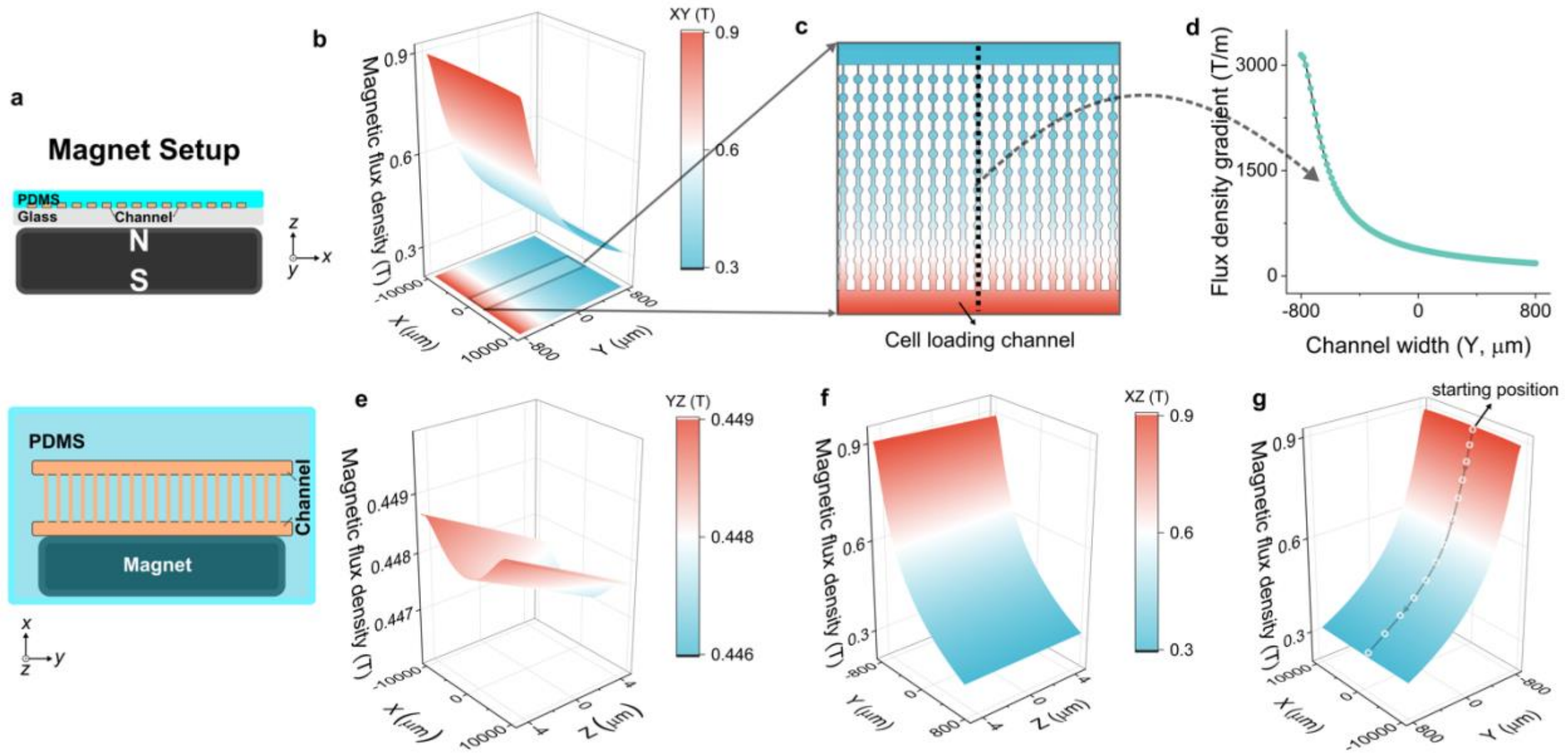

**Figure 2. Magnetic field configuration and flux density gradients in the Stiff-FCS device.** (a) Schematic of the Stiff-FCS platform illustrating the relative positioning of the permanent magnet and the microfluidic channel array. A rectangular N52 neodymium magnet with dimensions 50.8 × 6.35 × 6.35 mm (length × width × height) is placed beneath the glass substrate supporting the PDMS microchannels. (b) Simulated distribution of the magnetic flux density of the magnet setup in the x-y plane (z = 0). (c) Cross-sectional view of the microchannel array overlaid on the magnetic field map, with the highest magnetic flux density localized near the bottom surface of the device, corresponding to the cell loading channel. (d) Distributions of magnetic flux density gradient in the microchannel. A magnetic flux density of 3146 T/m in the x-y plane (z = 0) was located near the edge of the loading channel. (e) Simulated distribution of the magnetic flux density of the magnet setup in the x-z plane (y = 0). (f) Simulated distribution of the magnetic flux density of the magnet setup in the y-z plane (x = 0). (g) Calculated migration of diamagnetic particles toward regions of minimal magnetic field.

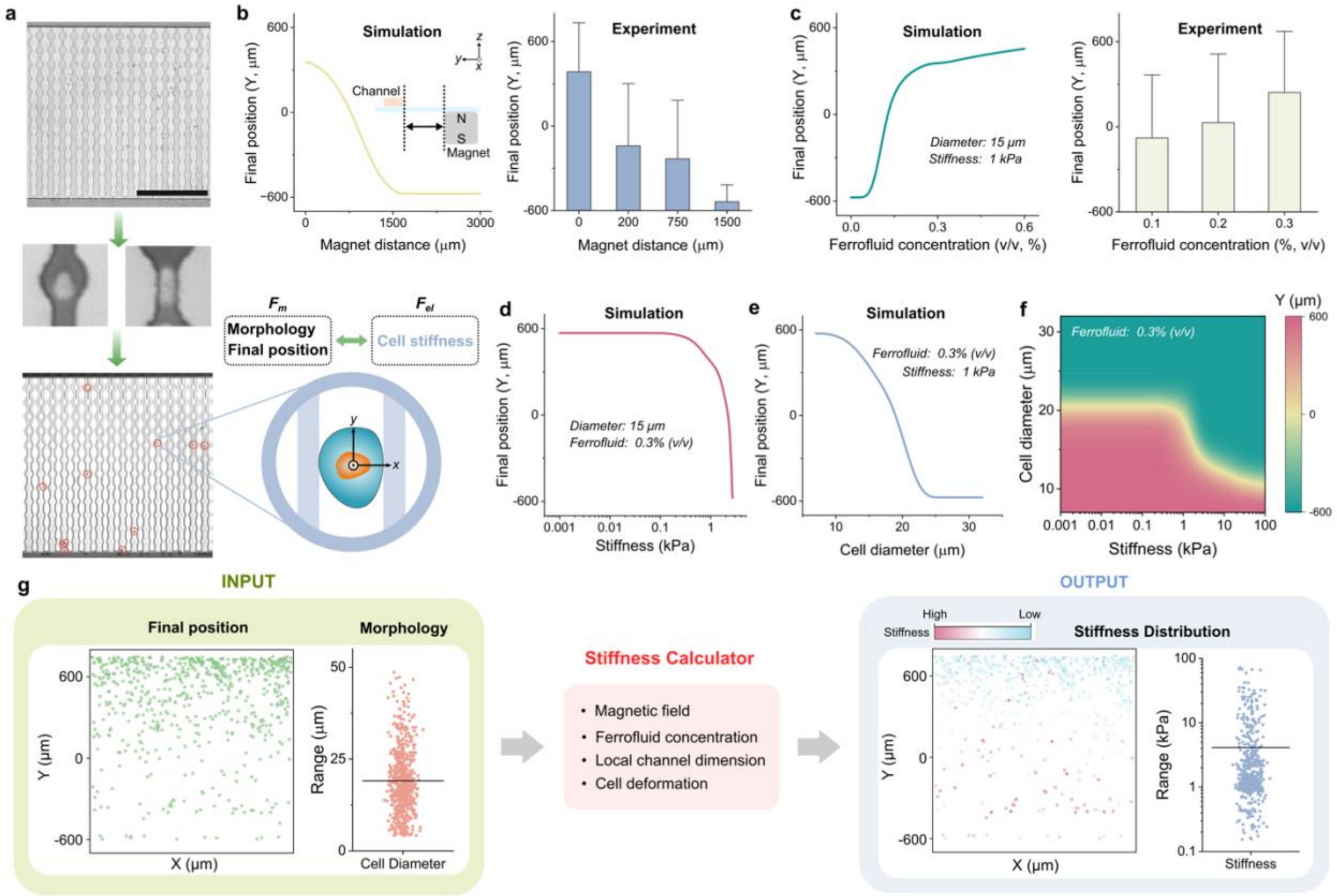

**Figure 3. Calibration of the Stiff-FCS method for stiffness-based cell sorting and stiffness calculation.** **(a)** Bright-field images of cell distributions in the Stiff-FCS device prior to magnetic field application (top), during lateral migration under magnetic buoyancy forcing (middle), and after sorting (bottom). Under steady state conditions, cells occupy distinct lateral positions where the magnetic buoyancy force $F_{mag}$ is balanced by an elastic compression force $\mathbf{F_{el}}$ arising from confinement-induced deformation. Cell stiffness is inferred from measured cell morphology, including equivalent diameter and deformation, together with the final lateral position that reflects the local magnetic buoyancy force. **(b)** Simulated (left) and experimental (right) final lateral positions Y as a function of magnet-channel distance, which controls the magnetic field strength. Experimentally measured final positions of 384.5 ± 348.4 μm (n = 314), −140.7 ± 442.2 μm (n = 510), −232.7 ± 416.1 μm (n = 877), and −537.6 ± 120.9 μm (n = 278) correspond to magnet-channel distances of 0, 200, 750, and 1500 μm, respectively (mean ± s.d.). Cell diameter is 15 μm, and ferrofluid concentration is 0.3% (v/v). **(c)** Simulated (left) and experimental (right) final lateral positions Y as a function of ferrofluid concentration. Experimentally measured final positions of −79.3 ± 445.1 μm (n = 620), 29.3 ± 484.2 μm (n = 589), and −242.7 ± 430.7 μm (n = 457) correspond to ferrofluid concentrations of 0.1, 0.2, and 0.3% (v/v), respectively (mean ± s.d.). Cell diameter is 15 μm. **(d)** Simulated dependence of final position Y on cell stiffness for a fixed cell diameter of 15 μm and ferrofluid concentration of 0.3% (v/v). **(e)** Simulated dependence of final lateral position Y on cell diameter for a fixed stiffness of 1 kPa and ferrofluid concentration of 0.3% (v/v). **(f)** Simulated final position Y as a function of cell stiffness and diameter, illustrating the operating regime in which stiffness can be uniquely inferred under fixed ferrofluid concentration. **(g)** Schematic of the stiffness calculation workflow. Measured cell final position and morphology are used as inputs to an inverse mechanical model based on steady-state force balance, incorporating local magnetic field strength, ferrofluid concentration, channel geometry, and confinement-induced deformation to compute single-cell stiffness distributions. Experimental validation was performed using the human non-small cell lung cancer cell line H1299.

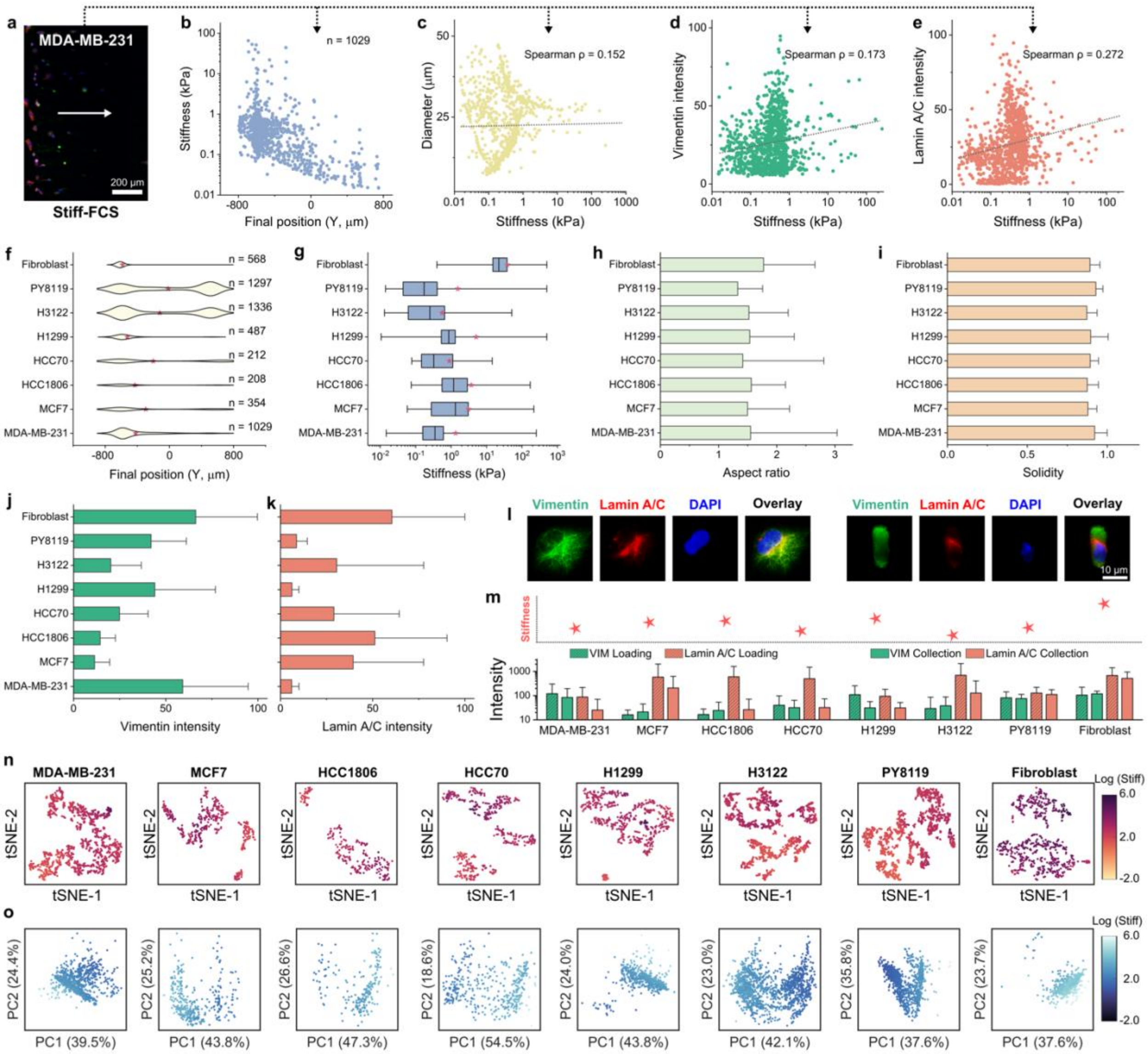

**Figure 4. Single-cell stiffness phenotyping and multivariate analysis enabled by Stiff-FCS**. **(a)** Immunofluorescence image of MDA-MB-231 human breast cells in the Stiff-FCS device after stiffness-based sorting. Three channels, including Vimentin (green), Lamin A/C (red), and DAPI (blue) are shown. The ferrofluid concentration is 0.3% (v/v). **(b)** Single-cell stiffness as a function of final position Y for MDA-MB-231 cells (n = 1029). (c) Relationship between single-cell stiffness and equivalent cell diameter. (Spearman ρ = 0.152, P < 0.001) (d) Relationship between single-cell stiffness and Vimentin intensity. (Spearman ρ = 0.173, P < 0.001). (e) Relationship between single-cell stiffness and Lamin A/C intensity (Spearman ρ = 0.272, P < 0.001). (f–k) Cross-cell-line comparison of final lateral position (f), single-cell stiffness (g), mean aspect ratio (h), mean solidity (i), mean Vimentin fluorescence intensity (j), and mean lamin A/C fluorescence intensity (k) for human breast cancer cell lines (MDA-MB-231, MCF7, HCC1806, HCC70), human non-small cell lung cancer cell lines (H1299, H3122), mouse breast cancer cells (PY8119), and human fibroblasts. Cell numbers are 1029, 354, 208, 212, 487, 1336, 1297, and 568, respectively. Box plots indicate the median, interquartile range ($25^{th}$-$75^{th}$ percentiles), and whiskers extending to 1.5 times the interquartile range. (l) Representative immunofluorescence images of MDA-MB-231 cells collected from the loading channel (near the magnet) and the collection channel (far from the magnet). (m) Mean

fluorescence intensity of Vimentin (green) and Lamin A/C (red) of cells collected from loading channel and the collection channel. Red stars indicate the mean single-cell stiffness of the corresponding cell lines and are shown as a visual reference. (n) t-SNE visualization of single-cell measurements incorporating final lateral position, equivalent diameter, stiffness, aspect ratio, solidity, Vimentin intensity, and Lamin A/C intensity for each cell line. Points are colored by log-transformed stiffness. (o) Principal component analysis of the same single-cell feature set for each cell line, with points colored by log-transformed stiffness and the percentage of variance explained by each principal component indicated.

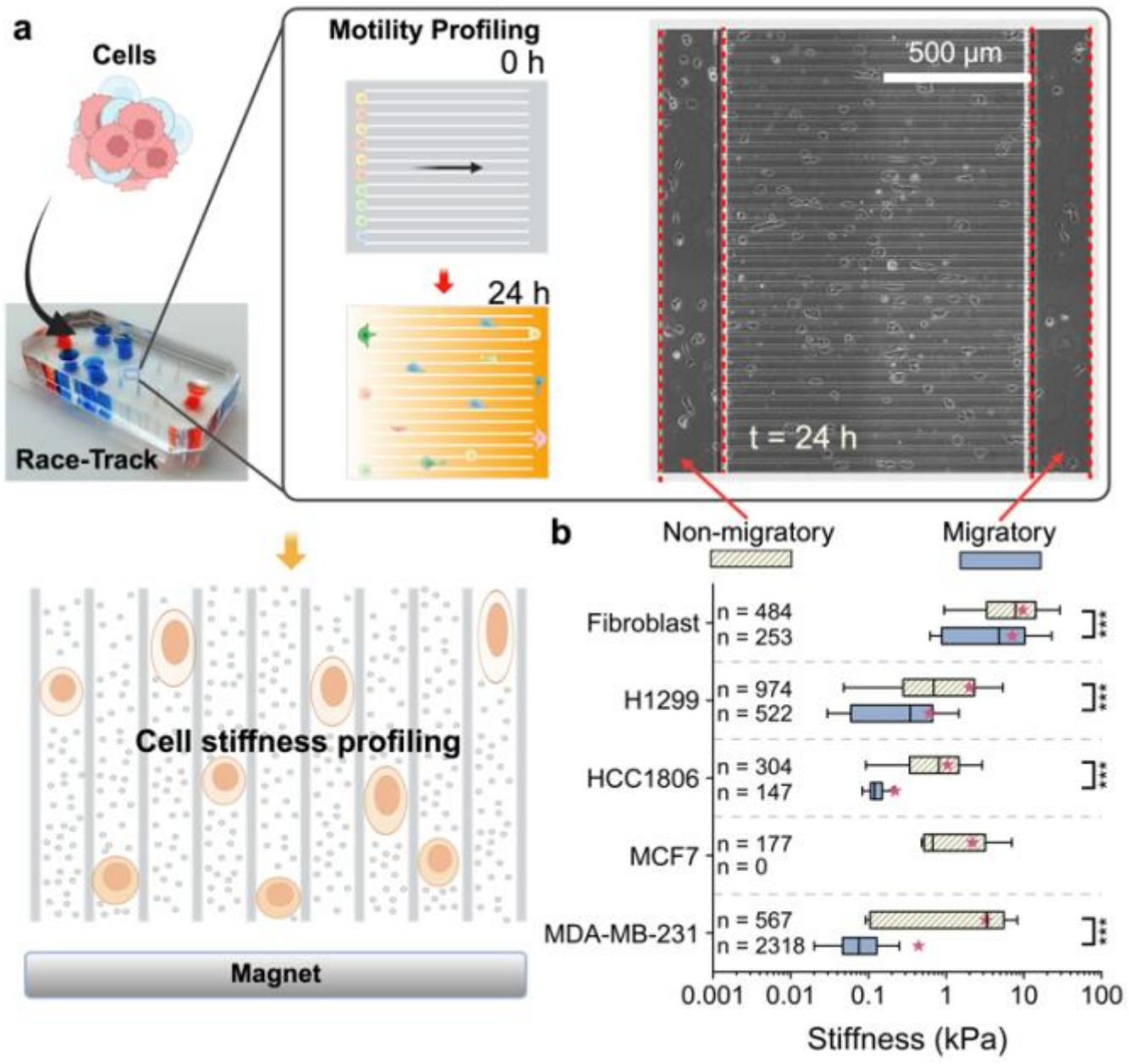

**Figure 5. Association between cell migratory behavior and single-cell stiffness measured by Stiff-FCS.** (a) Schematic illustration of the workflow for migration-based cell sorting followed by stiffness profiling. Cells were first introduced into the Race-Track device and subjected to chemotactic cues (10% FBS) to profile motility. Based on directional migration over 24 h, cells were classified as migratory or non-migratory and subsequently collected for stiffness measurement using Stiff-FCS. Created with BioRender.com released under a Creative Commons Attribution-NonCommercial-NoDerivs 4.0 International license. (b) Single-cell stiffness distributions of migratory and non-migratory subpopulations for fibroblasts, human non-small cell lung cancer cells (H1299), human breast cancer cells (HCC1806, MCF7, MDA-MB-231). Box plots indicate the median, interquartile range ($25^{th}$-$75^{th}$ percentiles), and whiskers extending to 1.5 times the interquartile range. Star symbols denote the mean stiffness for each cell line. n = 3 independent experiments using different devices. Statistical significance is indicated as *$P$ < 0.05, **$P$ < 0.01, and ***P < 0.001.

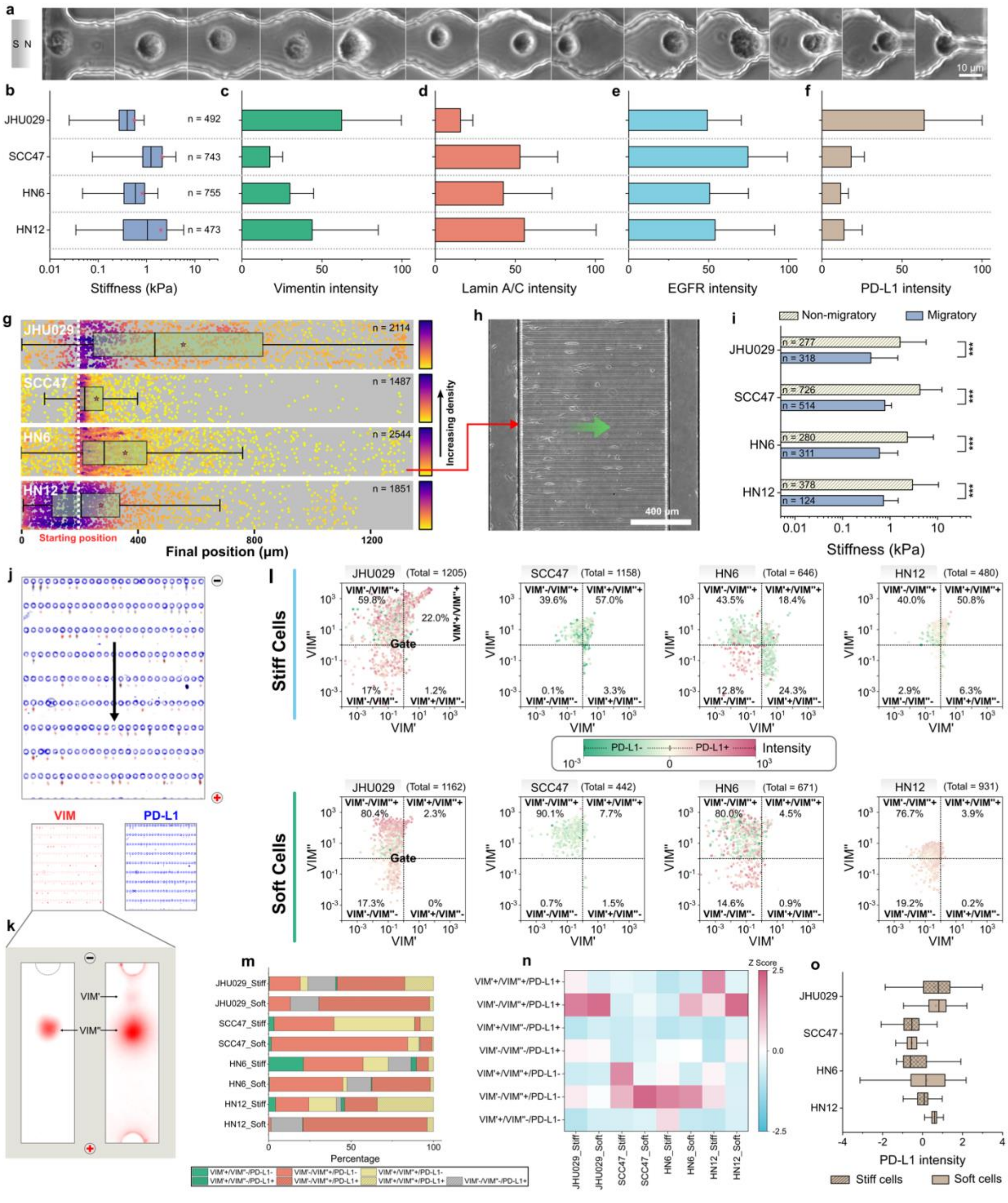

**Figure 6. Mechanical state-resolved molecular and migratory heterogeneity in human HNC cells.** (a) Representative bright-field image sequence showing deformation of individual JHU029 cells during stiffness profiling in the Stiff-FCS device. Cells move from left to right. Ferrofluid concentration is 0.3% (v/v). (b-f) Single-cell stiffness distributions (b) and corresponding mean fluorescence intensities of Vimentin (c), Lamin A/C (d), EGFR (e), and PD-L1 (f) for human HNC cell lines (JHU029, SCC47, HN6, and HN12). Star symbols denote mean stiffness values. (g) Migration distance profiles of HNC cell lines measured using the Race-Track assay after 24 h under a chemoattractant gradient (10% FBS). Star symbols

denote mean migration distances. (h) Representative image of the Race-Track device used for chemotaxis-based migration profiling. (i) Single-cell stiffness distributions of migratory and non-migratory subpopulations following migration-based separation for each HNC cell line. Statistical significance is indicated as $*P < 0.05$, $**P < 0.01$, and $***P < 0.001$. (j) Schematic illustration of stiffness-resolved single-cell western blot (scWB) for proteoform analysis of Vimentin and PD-L1 in stiff and soft cell subpopulations. (k) Fluorescence micrographs of single-cell western blots of JHU029 cells showing Vimentin signal. (l) Representative scWB scatter plots showing Vimentin proteoform (VIM' and VIM'') distributions for stiff and soft cells across HNC cell lines. Each point represents a single cell. Protein expression is considered negative when signal intensity is below 1 (arbitrary units). Point color indicates PD-L1 intensity. (m) Relative abundance of Vimentin and PD-L1 proteoform combinations in stiffness-resolved subpopulations. (n) Heatmap summarizing z-score normalized enrichment of Vimentin and PD-L1 proteoforms across stiff and soft subpopulations for each HNC cell line. (o) Box plots of PD-L1 fluorescence intensity for stiffness-resolved subpopulations. Boxes indicate the median and interquartile range ($25^{th}$-$75^{th}$ percentiles), with whiskers extending to 1.5 times the interquartile range. Squares denote mean values.

**Supplementary Information**

# Stiff-FCS: Single-Cell Stiffness Profiling With Integrated Molecular and Functional Analysis

Yuhao Zhang[a], Luyao Zhao[g], Zhengfu Huang[d], Kenan Song[d], Xianqiao Wang[d], Xianyan Chen[e], Jin Xie[g], Yiping Zhao[f], He Li[a], Leidong Mao[c], Yong Teng*[h,i], Yang Liu*[a,b]

[a]School of Chemical, Materials and Biomedical Engineering, College of Engineering, The University of Georgia, Athens, Georgia 30602, USA

[b]Institute of Bioinformatics, The University of Georgia, Athens, Georgia 30602, USA

[c]School of Electrical and Computer Engineering, College of Engineering, The University of Georgia, Athens, Georgia 30602, USA

[d]School of Environmental, Civil, Agricultural, and Mechanical Engineering, The University of Georgia, Athens, Georgia 30602, USA

[e]College of Public Health, The University of Georgia, Athens, Georgia 30602, USA

[f]Department of Physics and Astronomy, The University of Georgia, Athens, Georgia 30602, USA

[g]Department of Chemistry, The University of Georgia, Athens, Georgia, 30602, USA

[h]Department of Hematology and Medical Oncology, Winship Cancer Institute, Emory University, Atlanta, Georgia 30322, USA

[i]Wallace H. Coulter Department of Biomedical Engineering, Georgia Institute of Technology & Emory University, Atlanta, GA 30322, USA

*Email: Yang Liu (liuy@uga.edu); Yong Teng (yong.teng@emory.edu)

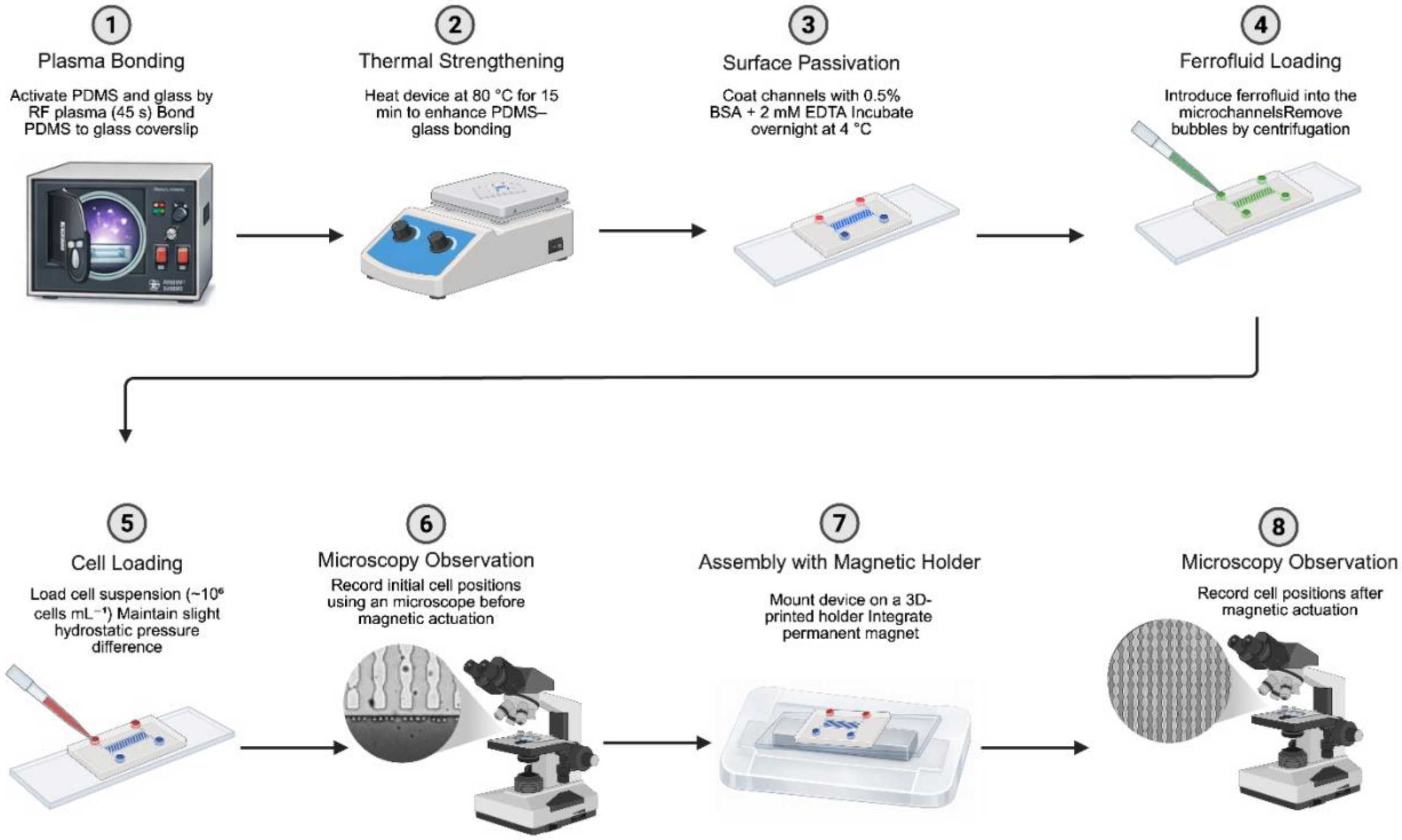

**Figure S1. Experimental workflow of Stiff-FCS for ferrohydrodynamic cell manipulation and stiffness profiling.** Schematic illustration of the experimental procedure. (1) PDMS devices were plasma-bonded to glass coverslips. (2) Bonding was thermally strengthened at 80 °C for 15 min. (3) Microchannels were passivated with 0.5% BSA and 2 mM EDTA and incubated at 4 °C overnight to minimize nonspecific adhesion. (4) Ferrofluid was introduced into the microchannels, and air bubbles were removed prior to experiments. (5) Cell suspensions (~$10^6$ cells $mL^{-1}$) were loaded into the device under a slight hydrostatic pressure difference. (6) Initial cell positions were recorded by optical microscopy before magnetic actuation. (7) The device was mounted onto a magnetic holder integrating a permanent magnet to generate a spatial magnetic field. (8) Cell displacement and positioning were recorded under magnetic actuation for subsequent stiffness analysis. Created with BioRender.com released under a Creative Commons Attribution-NonCommercial-NoDerivs 4.0 International license.

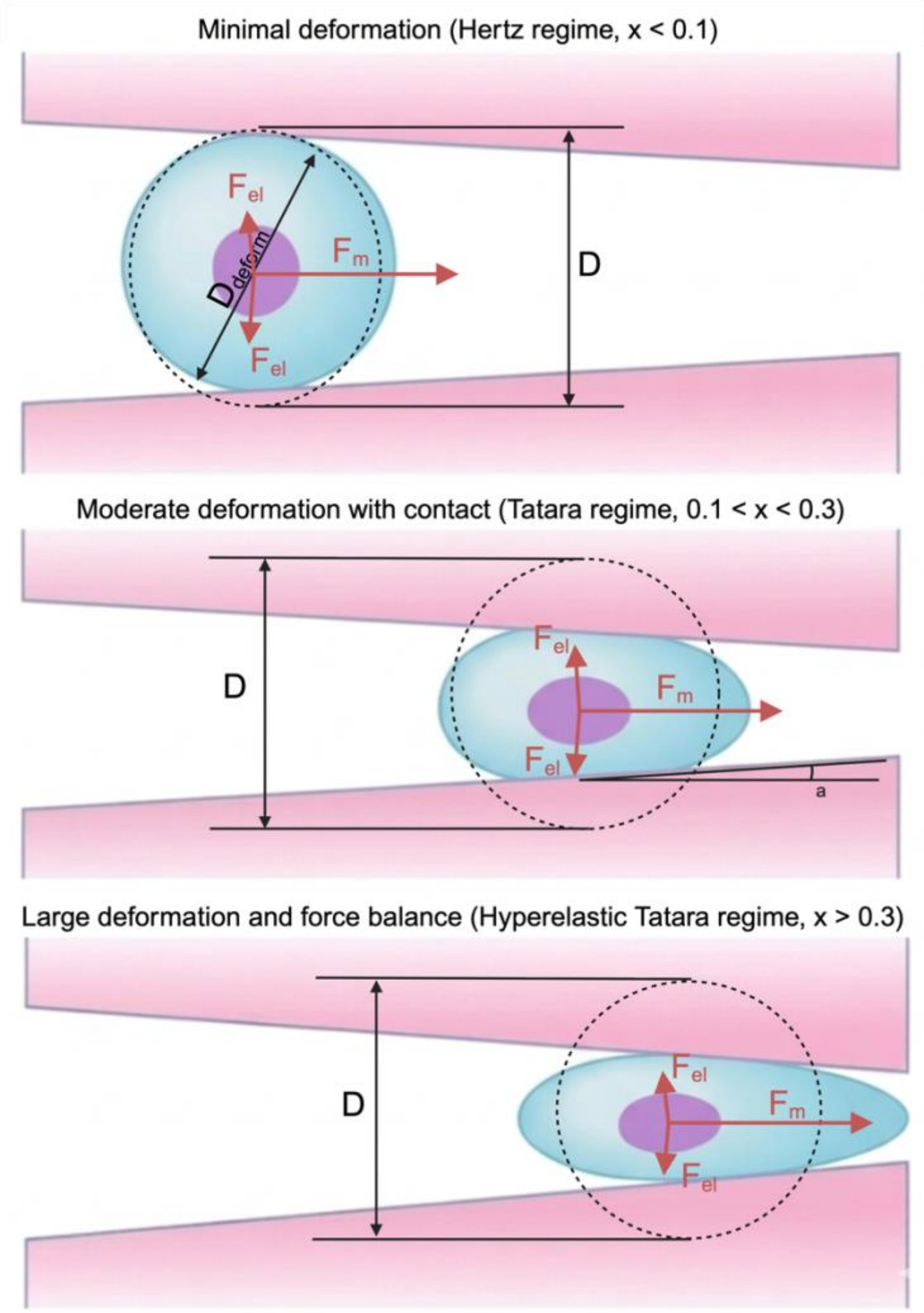

**Figure S2. Elastic compression of cells under geometric confinement in a tapered microchannel.** Schematic of cell deformation and force balance as a cell traverses a narrowing channel of width $D$, smaller than its undeformed diameter. Geometric confinement induces symmetric compression by rigid channel walls, generating an elastic compression force ($F_{\mathrm{el}}$) that opposes external magnetic force ($F_{\mathrm{m}}$). The deformation ratio is defined as $x = (D_{\mathrm{cell}} - D_{\mathrm{deform}})/D_{\mathrm{cell}}$, where $D_{\mathrm{deform}}$ denotes the apparent cell diameter perpendicular to the compression direction. Three deformation regimes are shown. **Top**, minimal deformation ($x < 0.1$), where the cell remains approximately spherical and is described by the Hertz model. **Middle**, moderate deformation ($0.1 < x < 0.3$), where wall contact induces geometric flattening and the response is captured by the Tatara model. **Bottom**, large deformation ($x > 0.3$), where both geometric flattening and material nonlinearity become significant, and the behavior is described by the hyperelastic Tatara model. Created with BioRender.com released under a Creative Commons Attribution-NonCommercial-NoDerivs 4.0 International license.

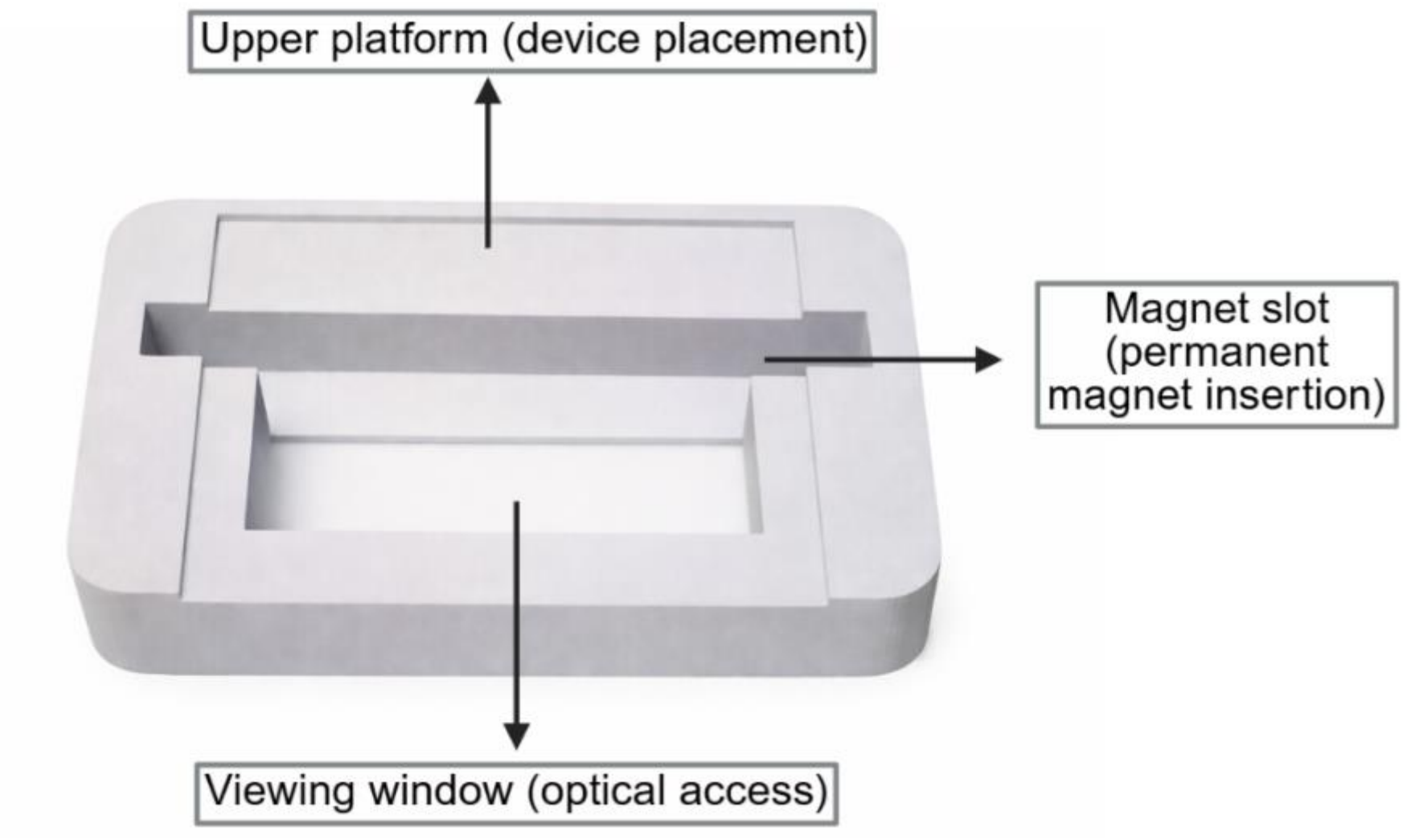

**Figure S3. Custom holder for magnet integration and device positioning.** The microfluidic device is mounted on a custom-designed holder featuring an upper platform for device placement, a central viewing window for real-time observation, and a recessed slot for permanent magnet insertion. The holder enables precise control of the magnet–device distance, thereby regulating the magnetic field applied to the screening region.

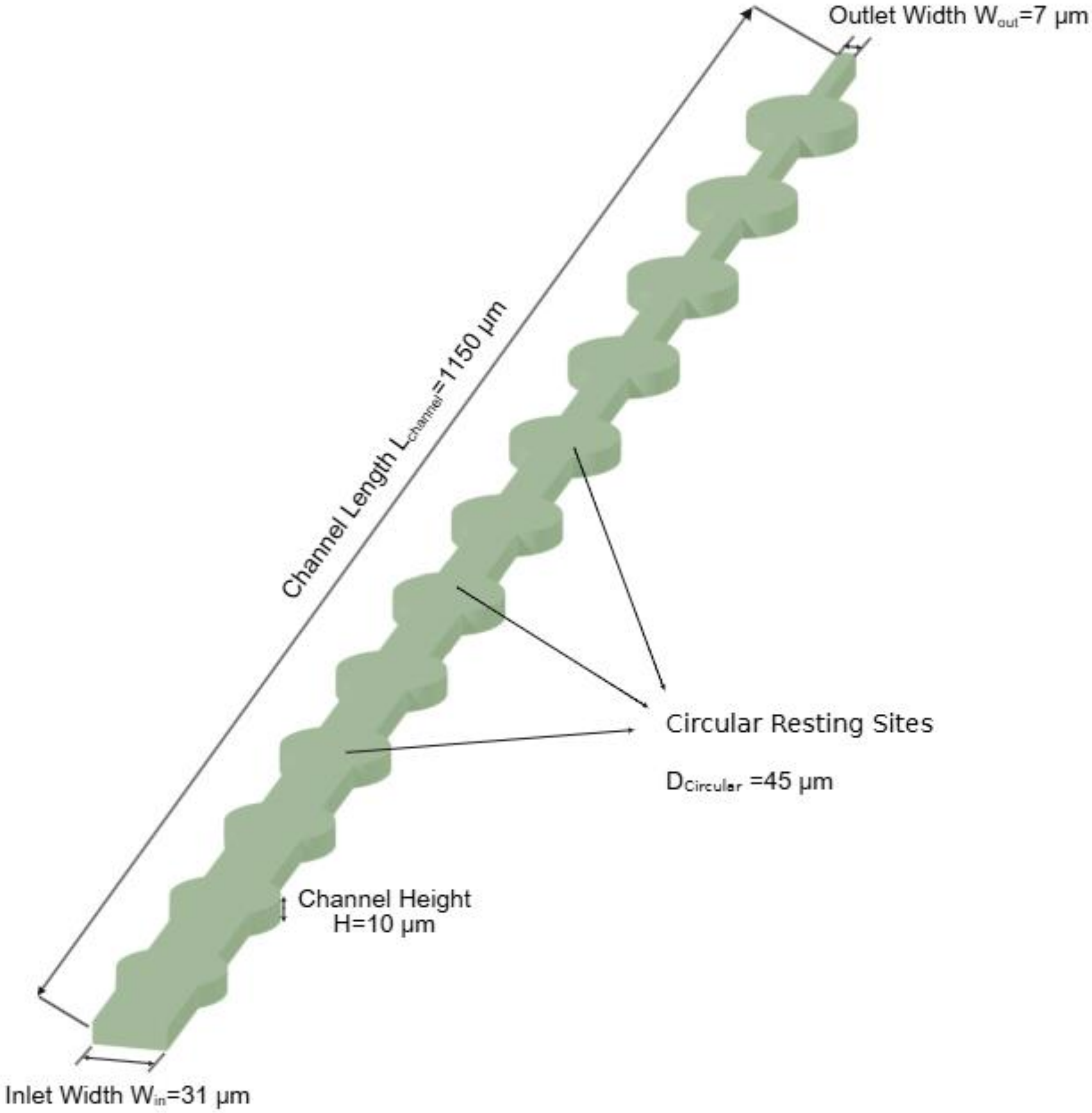

**Figure S4. Geometry of a single grading microchannel in the Stiff-FCS device.** Schematic illustration of a single grading microchannel composed of repeated constriction units with integrated circular resting sites. The channel width decreases monotonically from an inlet width of $W_{\text{in}} = 31\,\mu m$ to an outlet width of $W_{\text{out}} = 7\,\mu m$ over a total length of $L_{\text{channel}} = 1150\,\mu m$, establishing a continuous confinement gradient for deformability-based cell positioning. Circular resting sites (diameter $D_{\text{circular}} = 45\,\mu m$) are incorporated along the channel to facilitate cell retention and handling. The channel height is $H = 10\,\mu m$.

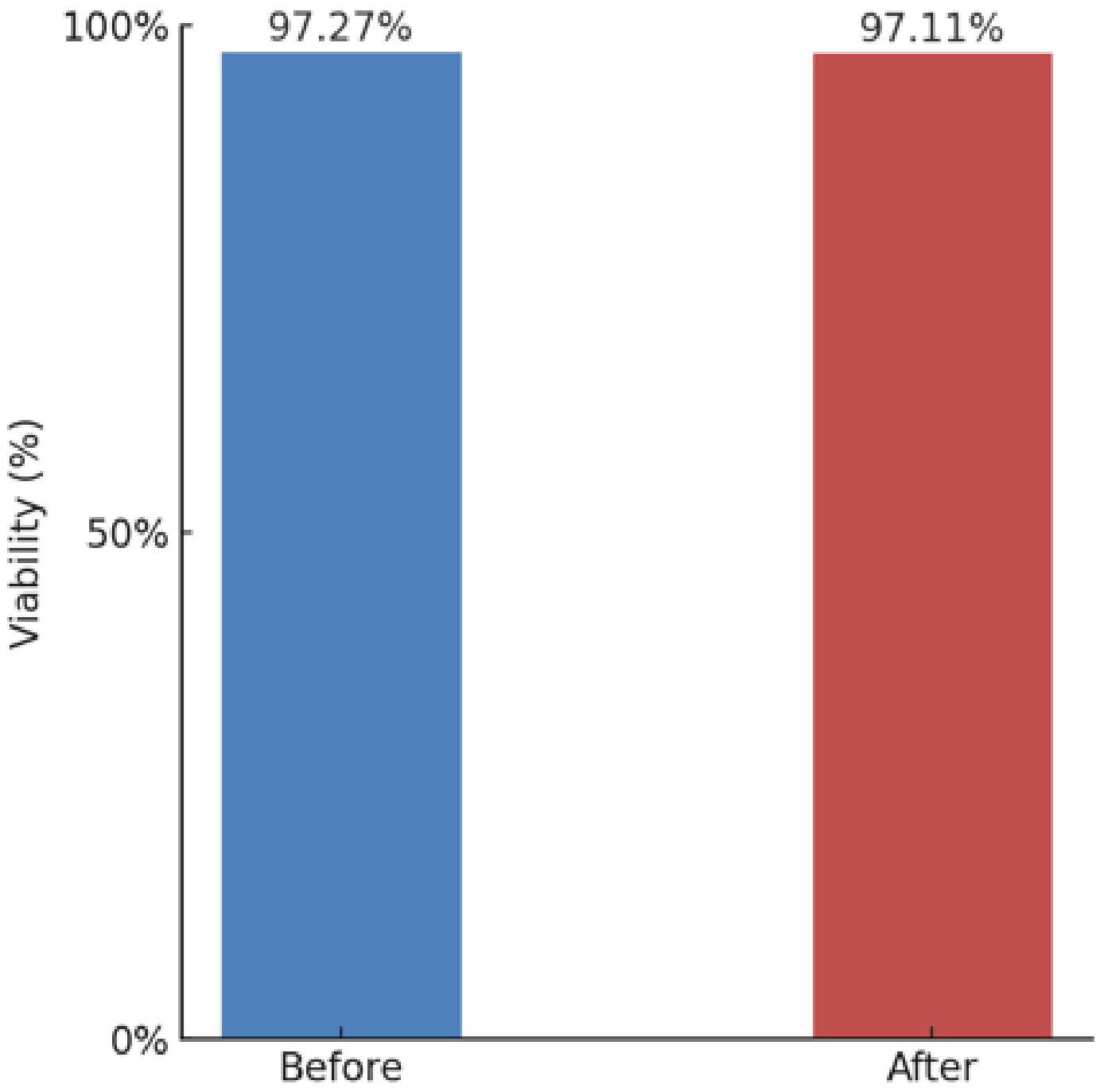

**Figure S5. Cell viability of MDA-MB-231 cells before and after Stiff-FCS processing.** Cell viability measured by Live/Dead staining remained high before and after microfluidic processing of MDA-MB-231 cells (97.27% before vs 97.11% after), indicating minimal impact of the Stiff-FCS platform on cell viability. Live cells were identified by Calcein AM fluorescence, and dead cells by ethidium homodimer-1 staining. Microchannels were collagen-coated and pre-filled with 0.3% (v/v) ferrofluid prior to cell loading.

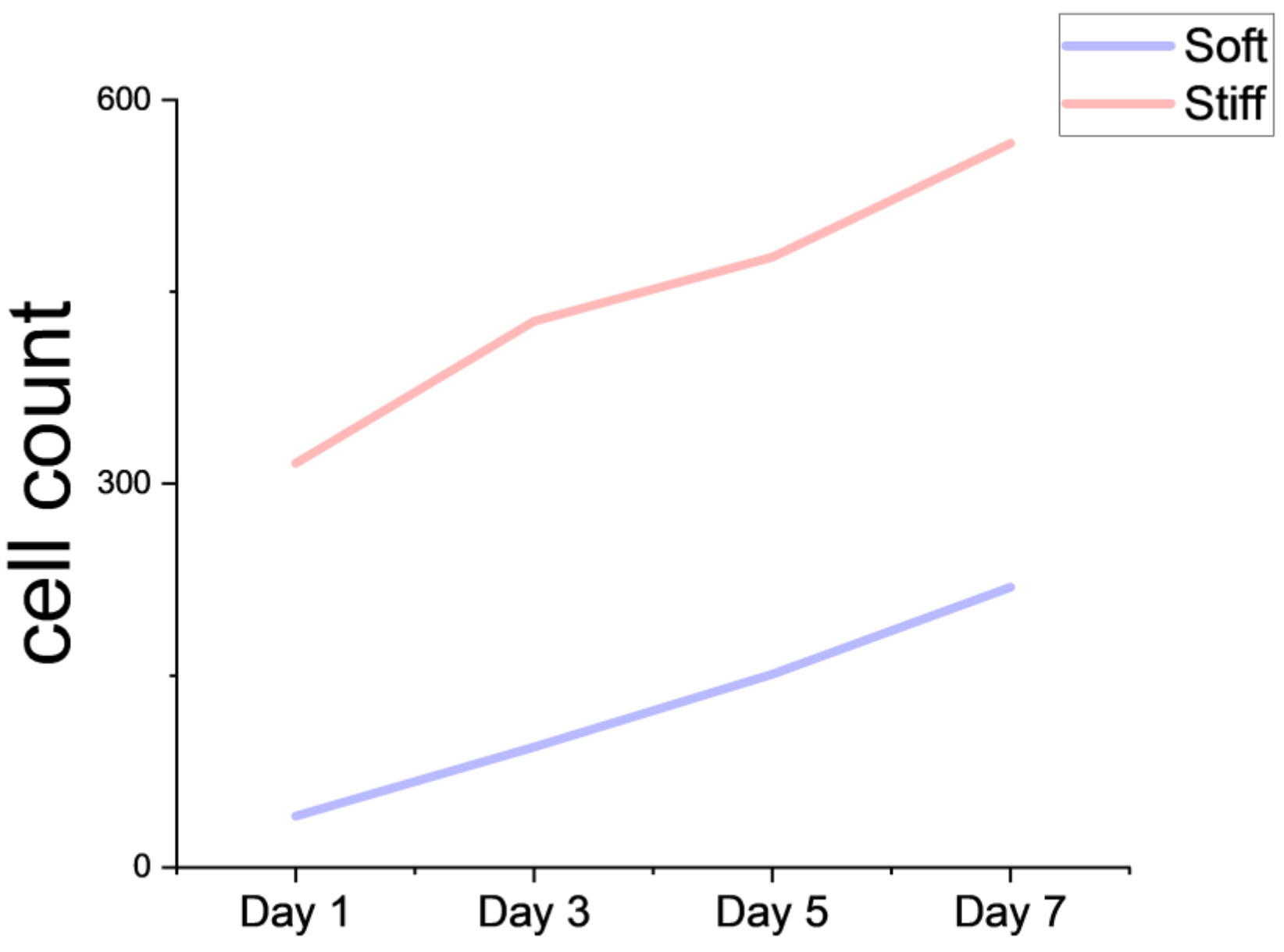

**Figure S6. Stiffness-resolved MDA-MB-231 cells retain proliferative capacity after Stiff-FCS separation.** Proliferation of stiffness-resolved MDA-MB-231 subpopulations after Stiff-FCS separation. Cells recovered from the loading channel were defined as stiff cells, and cells recovered from the collection channel were defined as soft cells. After washing to remove residual ferrofluid, equal volumes of each cell suspension were seeded into 96-well plates and cultured under standard conditions. Cell proliferation was monitored on days 1, 3, 5, and 7 after seeding, showing that both recovered subpopulations remained amenable to post-separation culture.

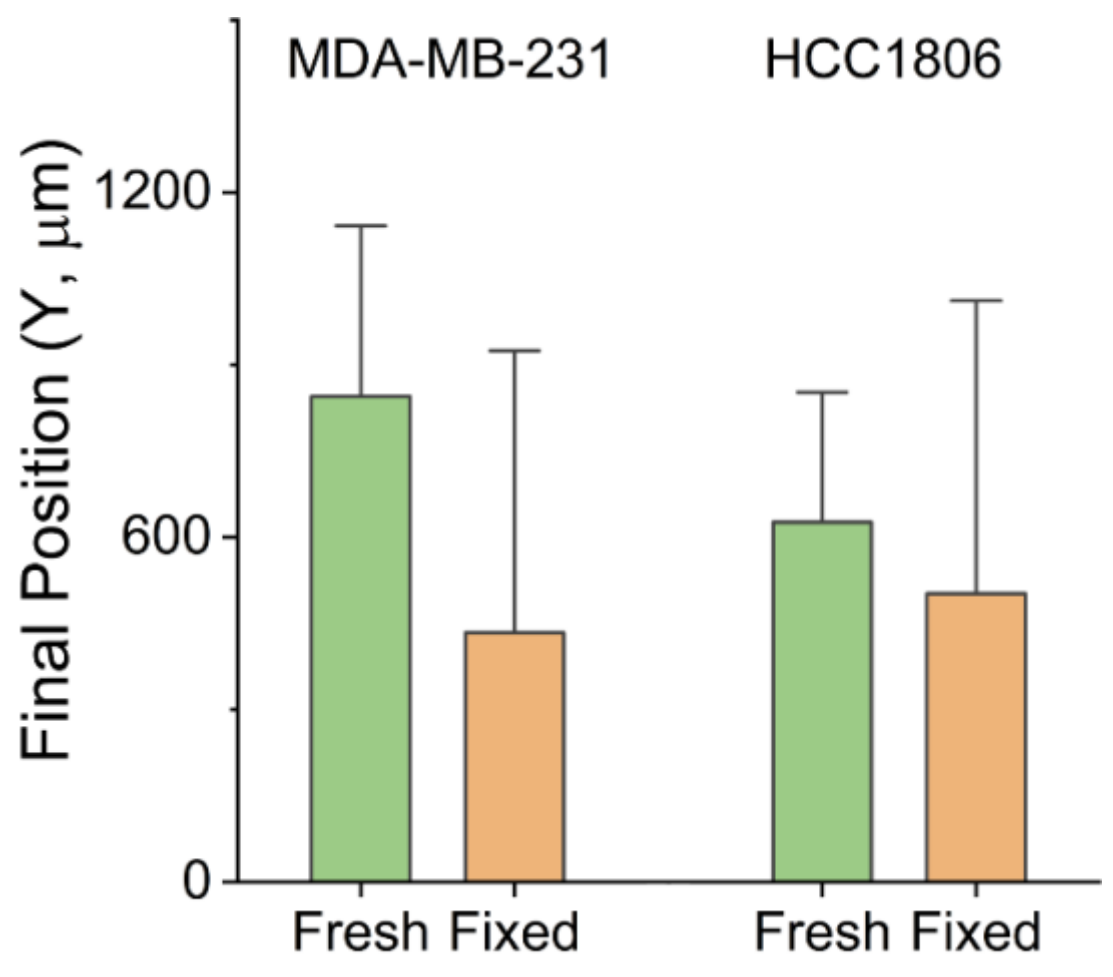

**Figure S7. PFA-fixed cells show reduced final positions in Stiff-FCS.** Final positions of fresh and PFA-fixed MDA-MB-231 and HCC1806 cells measured in Stiff-FCS. PFA-fixed cells tended to reach lower final positions than fresh controls in both cell lines, consistent with the expected behavior of mechanically stiffer cells under confinement-based sorting. The final positions were 844.77 ± 296.83 μm (fresh, n = 384) and 434.18 ±490.29 μm (fixed, n = 282) for MDA-MB-231, and 626.22 ± 225.87 μm (fresh, n = 654) and 501.90 ± 509.43 μm (fixed, n = 400) for HCC1806. Data are presented as mean ± s.d.

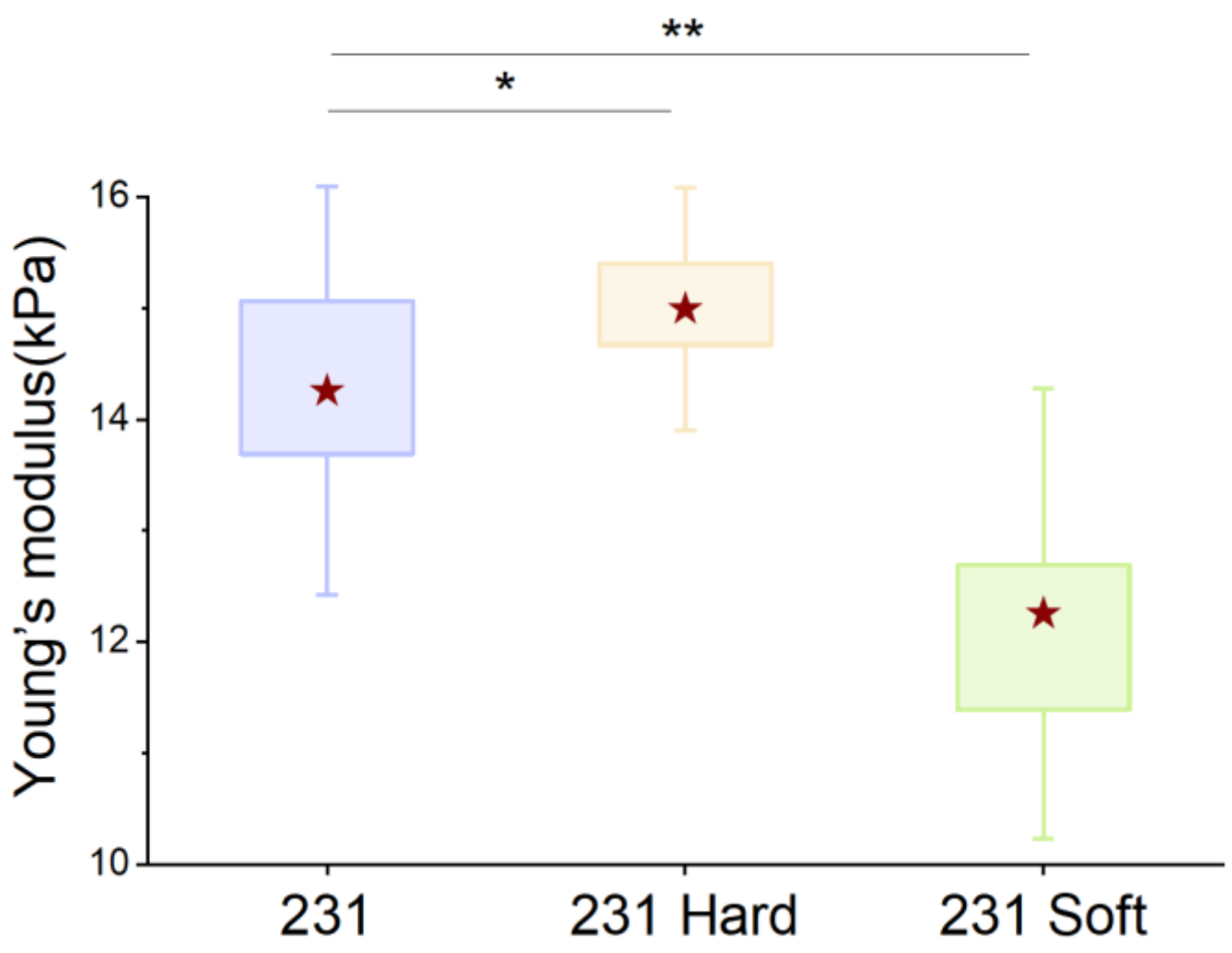

**Figure S8. AFM validation of the parental MDA-MB-231 population and Stiff-FCS-resolved subpopulations.** Apparent Young's modulus of parental MDA-MB-231 cells and the corresponding Hard and Soft subpopulations isolated by Stiff-FCS, measured by atomic force microscopy (AFM). Each point represents one cell; box plots show the median and interquartile range. Approximately 10 cells were analyzed per condition. Limited number due to the low-throughput nature of AFM measurements. Compared with the parental population, the Hard fraction displayed increased stiffness, whereas the soft fraction displayed reduced stiffness. *: $P < 0.05$, **: $P < 0.01$.

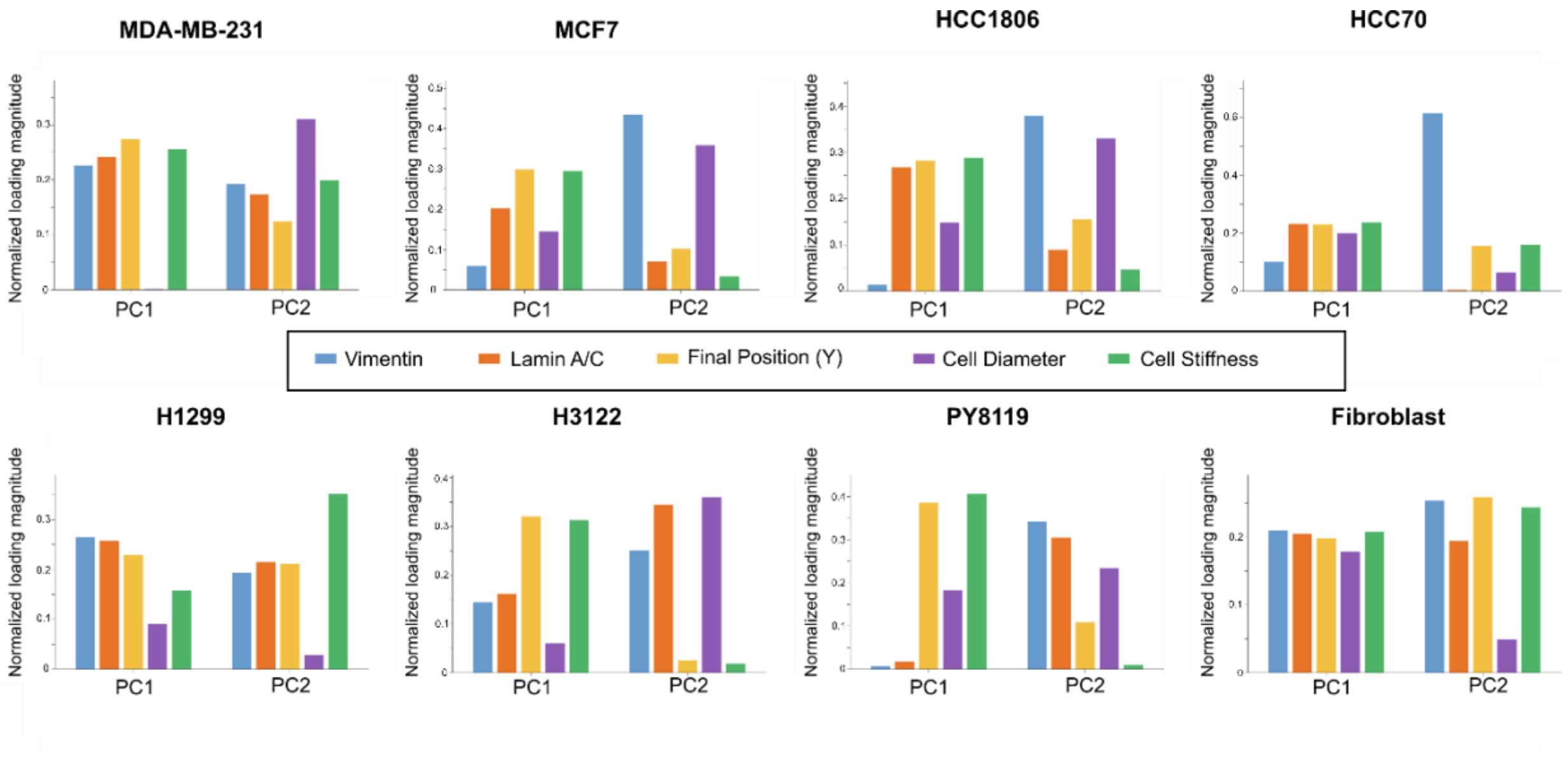

**Figure S9. Condition-specific principal component loadings reveal differential contributions of protein expression and biophysical features to PC1 and PC2.** PCA loading magnitudes for PC1 and PC2 across the tested cell lines. The relative contributions of Vimentin intensity, Lamin A/C intensity, final position (Y), cell diameter, cell stiffness, and other measured single-cell features varied across cell lines, indicating that the observed phenotypic heterogeneity cannot be explained by a single universal feature. Together, these results show that Stiff-FCS enables multidimensional phenotypic mapping of single cells.

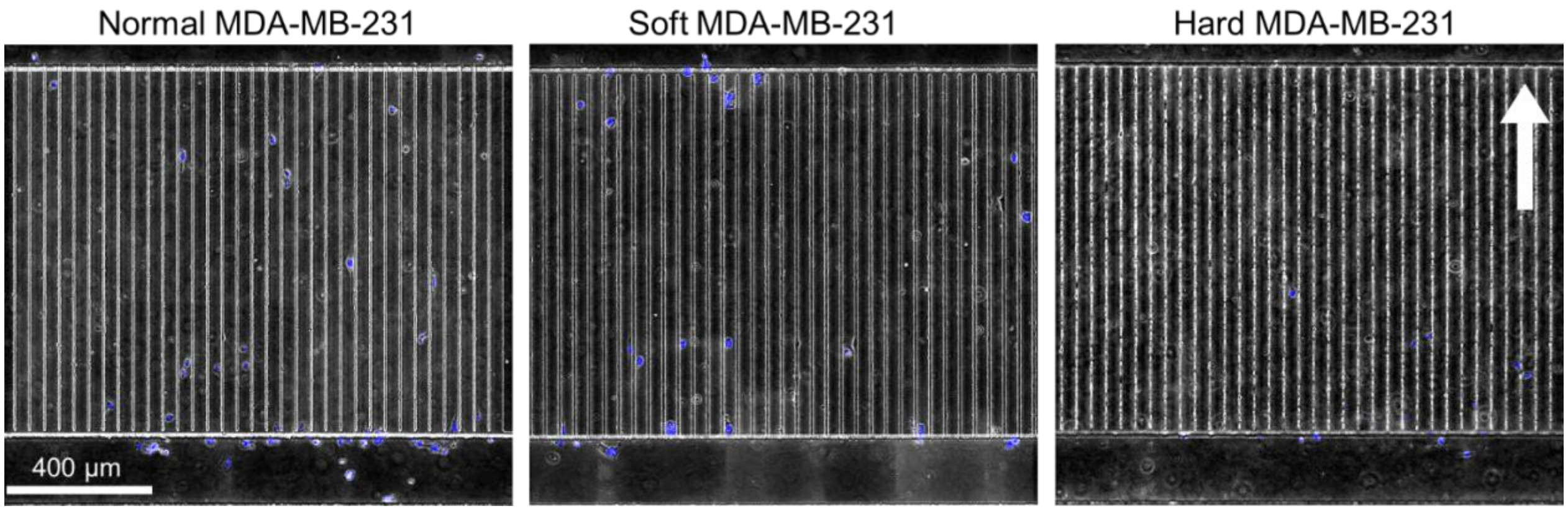

**Figure S10. Migration profiles of stiffness-resolved MDA-MB-231 cells.** Representative fluorescence images of MDA-MB-231 cells categorized into normal, soft, and stiff subpopulations following stiffness-based separation and subjected to migration assays. Nuclei were stained with DAPI (blue). Differences in spatial distribution and migration patterns are observed among the three groups. Scale bar, 400 µm.

**Table S1. Overview of representative microfluidic platforms for single-cell mechanical phenotyping.**

| Major method category | Technique name | Paper title | Contact level | Throughput | Cell viability | Cell retrieval | Downstream assay possible |
|---|---|---|---|---|---|---|---|
| **Category 1: Constriction-/filtration-based** | Stiffness based isolation in microfluidic flow chamber | Numerical simulations of cell flow and trapping within microfluidic channels for stiffness based cell isolation. [1] | Direct contact and squeezing through 10 µm openings | 10–100 cells/s (≈ 3.6 × $10^4$ cells/h) | 85–95 % (cells subjected to <10 nN force remain viable) | Yes (stiff cells retained, soft cells pass) | Yes (passed cells usable) |
| | Passive circulating cell sorting by deformability using a microfluidic gradual filter (Preira et al. 2013) | Passive circulating cell sorting by deformability using a microfluidic gradual filter. [2] | Repeated direct contact in a series of narrowing gaps | 5–20 cells/s | > 90 % | Partly (passed and trapped fractions defined) | Yes, for passed cells |
| | Stiffness based filtration platform with replaceable membrane | Biophysical properties of cells for cancer diagnosis.[3] | Direct contact through membrane pores | $10^3$–$10^5$ cells/s effective (with $10^7$–$10^8$ pores membrane) | 80–90 % | Yes (membrane replaced and fractions recovered) | Yes |
| | DLD + trapping barrier array | Microfluidic cytometric analysis of cancer cell transportability and invasiveness.[4] | Direct contact and dynamic squeezing through decreasing microgaps (15–4 µm) | ~$10^3$–$10^4$ cells/run; ~4500 cells per assay at 10 µL/min | Not quantitatively reported; , probably 100 % | No | Yes (on chip IF staining; correlation with AFM and biomarkers) |
| | Linearly tapered microfluidic channel for deformability-based characterization | Microfluidic channel for characterizing normal and breast cancer cells.[5] | Direct contact and strong deformation in tapered microchannel (30 → 5 µm) | Low–moderate (single-cell passage, ~0.1–1 cells/s effective) | No (not evaluated or reported) | No | No |
| | Integrated microfluidic impedance cytometry with constriction-induced deformation | High-Precision Microfluidic Impedance Cytometry Integrating Cellular Size and Stiffness Sensing for Enhanced Tumor Cell Detection.[6] | Direct contact via constriction-induced deformation during flow | ≈$10^3$–$10^4$ cells/s | >90% | Yes (cells exit the device and can be physically collected) | Yes (off-chip downstream assays are possible, though not demonstrated in this study) |
| | Pressure-controlled U-shaped microfluidic traps with elasticity measurement and deterministic printing | Continuous trapping, elasticity measuring and deterministic printing of single cells using arrayed microfluidic traps.[7] | Direct contact, localized confinement with mild deformation (aspiration channel) | ≈5 cells/min per chip (≈300 cells/h; measured: ~18 s per cell printing; whole cycle 6.18 ± 1.44 min for 16 cells) | Not explicitly quantified; cells remain viable based on successful trapping, deformation, release, and printing | Yes (single cells deterministically printed into wells) | Yes (single cells printed into standard well plates) |
| | Impedance–deformability cytometry (hydrodynamic | Microfluidic Impedance-Deformability Cytometry for Label- Free Single Neutrophil Mechanophenotyping.[8] | Noncontact hydraulic deformation (no constriction) | ≈1000 cells/min (~16 cells/s per chip) | Yes (membrane integrity and cell state evaluated | Yes (cells flow through intact) | No |
| | pinching + differential impedance) | | | | electrically; no % viability reported) | | |
| | Continuous-flow deformability-based CTC separation using microfluidic ratchets | Continuous Flow Deformability-Based Separation of Circulating Tumor Cells Using Microfluidic Ratchets.[9] | Direct contact with tapered microconstrictions; transient squeezing through funnel-shaped pores under oscillatory flow | Approx. 1 mL whole blood/h/device; typically two devices run in parallel | Yes; enriched UM-UC13 cells retained 99.1% viability relative to pre-enriched cells | Yes; separated cells are collected in suspension from the outlet reservoir | Yes; collected cells can be fixed, stained, cultured, and used for downstream molecular characterization |
| | Differential impedance-based constriction microfluidic cytometry for simultaneous deformability and electrical characterization | Characterizing Deformability and Electrical Impedance of Cancer Cells in a Microfluidic Device.[10] | Direct contact and squeezing through a 10 µm-wide constriction; electrical sensing via four pairs of coplanar electrodes | >1000 cells/min | Not directly reported | Not demonstrated explicitly | Potentially yes in principle, but not demonstrated in this paper; authors propose future integration with real-time sorting |
| | PDMS capillary-mimicking microchannel assay measuring entry time, transit velocity, and elongation index | Deformability study of breast cancer cells using microfluidics.[11] | Direct contact and squeezing through a 10 µm × 10 µm square microchannel, 150 µm long | Low; not explicitly reported (video-based single-cell analysis with high-speed imaging) | Not reported | Not demonstrated explicitly | Not demonstrated explicitly |
| | Cell separation using microfluidic funnel ratchets under oscillatory flow | Cell separation based on size and deformability using microfluidic funnel ratchets.[12] | Direct contact and squeezing through funnel-shaped constrictions arranged in a 2D array | Approximately 9000 cells/h under standard operating conditions | Yes; >95% of MLCs remained viable before and after sorting over tested oscillation pressures of 7–41 kPa | Yes; the paper states that irreversibly transported cells can be completely cleared from the separation microstructure before introducing a new batch | Not demonstrated |
| | Microbarrier-array microfluidic chip with decreasing gaps for deformability-based separation of cancer cell subpopulations | Microfluidic Analysis for Separating and Measuring the Deformability of Cancer Cell Subpopulations.[13] | Direct contact and continuous squeezing/deformation while cells pass through arrays of blocking columns with gaps decreasing from 15 to 7 μm | 1 mL/h sample injection rate reported | Not reported | Yes; cells flowing out of the chip were collected, and cells blocked in the chip were also analyzed on-chip | Yes; collected cells were used for AFM, flow cytometry, and western blotting, and blocked cells were used for immunofluorescence analysis |
| | Tandem mechanical sorting with a deformability-based MS-Chip followed by a low-adhesion HCA-Chip | Microfluidic Tandem Mechanical Sorting System for Enhanced Cancer Stem Cell Isolation and Ingredient Screening.[14] | Direct mechanical interaction with sequentially variable microbarriers in the MS-Chip, followed by matrix-interacting fluid-mixing microchannels in the BME-coated HCA-Chip | Not directly reported as cells/s or cells/h; the MS-Chip was described as handling separation of millions of cells in less than 30 min, and sorting was operated at 60 μL/min in MS-Chip and 0.3 mL/h in HCA-Chip | Yes; CFSE/PI staining, trypan blue exclusion, and CCK8 assays showed no significant effect on viability or mortality after sorting | Yes; deformable cells were collected from the MS-Chip outlet, then reloaded into the HCA-Chip, and flexible/low-adhesive cells were collected from the final outlet for subsequent experiments | Yes; sorted cells were used for migration/invasion assays, wound healing, western blotting, colony formation, spheroid formation, flow cytometry, drug-resistance assays, xenograft studies, and compound screening |
| | PDMS confined-migration microchannel array with decreasing channel widths for analysis of steric hindrance during 3D migration | Migration in Confined 3D Environments Is Determined by a Combination of Adhesiveness, Nuclear Volume, Contractility, and Cell Stiffness.[15] | Direct physical confinement in linear PDMS microchannels; cells actively migrate through channels with widths decreasing from 11.2 to 1.7 μm and height 3.7 μm | Not reported | Not reported | Not demonstrated | Not demonstrated |
| | Multiconstriction microfluidic channel analysis using velocity profiles across repeated deformation–relaxation stages | Single-Cell Mechanical Characteristics Analyzed by Multiconstriction Microfluidic Channels.[16] | Direct contact and squeezing through repeated 8 µm × 8 µm constrictions separated by relaxation regions | Not reported | Cell viability before experiments was reported as 100% by trypan blue, but post-assay viability was not reported | Not demonstrated | Not demonstrated |
| | Mechanical separation chip (MS-chip) using artificial microbarriers | Microfluidics separation reveals the stem-cell–like deformability of tumor-initiating cells.[17] | Direct contact and squeezing through post gaps; cells interact with microbarriers whose gaps decrease from 15 µm to 7 µm; post height is ~13 µm | For the 75 × 25 mm chip, a typical run lasted ~15 min at ~1 mL/h; for the 75 × 50 mm chip, ~0.5 million cells could be processed in ~15 min at 2 mL/h | Flexible and stiff cells showed similar viability after separation; four independent groups were 89%, 89%, 84%, 75% for flexible cells and 84%, 85%, 83%, 78% for stiff cells | Yes. Flexible cells were collected from the outlet; stiff cells remained in the chip and were recovered by back-flushing to the inlet | Yes. Separated cells were used for gene-expression analysis, flow cytometry, and mammosphere formation assay |
| | Pressure-based funnel constriction microfluidic assay for RBC deformability | Microfluidic biomechanical assay for red blood cells parasitized by Plasmodium falciparum.[18] | Direct contact and squeezing through funnel-shaped constrictions; cells are deformed through serial funnel openings of 8, 7, 6, 5, 4, 3, 2.5, 2, 1.5, and 1 µm; channel height 3.7 µm | Not reported in cells/s; cells were measured individually, with at least three repeated tests per cell in the same funnel. | Not directly quantified, but measurements were considered valid only when cells showed no visible signs of damage and no permanent shape change | Yes, outlet flushing of measured cell is described, but this is for removal after measurement, not enrichment/sorting recovery | Potentially yes for downstream study, because the authors state the method can be used to study infected-cell properties and drug effects, but no downstream biological assay was directly demonstrated |
| | Microfluidic biophysical flow cytometry using parallel capillary-like microchannels | Analyzing cell mechanics in hematologic diseases with microfluidic biophysical flow cytometry.[19] | Direct physical confinement and deformation in capillary-like microchannels; cells enter 64 parallel microchannels that are 5.89 ± 0.08 µm wide, 13.3 µm tall, and 130 µm long | ~50–100 cells/min | Patient leukemia cell viability before assay was >95% by trypan blue exclusion | No recovery for downstream use demonstrated; after passing the device, cells drained to a waste container | Not directly demonstrated as a post-retrieval assay platform, although the paper notes optical imaging could in principle integrate multiparameter readouts such as fluorescence |
| | ParsortixTM cell separation system | A novel microfluidic platform for size and deformability based separation and the subsequent molecular characterization of viable circulating tumor cells.[20] | Direct physical confinement/contact within a stepped microstructure; cells are retained at a final separation gap of 10 µm and harvested by reverse flow | Not reported directly as cells/s or mL/h in this paper | 99% viable after processing in spiking experiments and after harvesting from patient samples | Yes; captured cells were subsequently harvested from the device, with average harvest yields of 54–69% of captured cells | Yes; downstream analyses demonstrated by mRNA characterization (RT-PCR) and array-based comparative genomic hybridization (aCGH) |
| | Slanted ridge deformability sorter (Islam et al. 2017; 2018; Wang et al. 2013) | Microfluidic sorting of cells by viability based on differences in cell stiffness.[21] | Limited contact, flow-guided deflection | 500–2 000 cells/s | > 95 % | Yes (sorted into outlet streams) | Yes |
| | Characterization and sorting of cells based on stiffness contrast (Sajeesh et al. 2016) | Characterization and sorting of cells based on stiffness contrast in a microfluidic channel.[22] | Light contact with walls | 50–200 cells/s | 90–95 % | Yes | Yes |
| | Multistage DLD + cone-channel stiffness sorting | Multistage microfluidic cell sorting method and chip based on size and stiffness.[23] | Direct physical contact (DLD guidance + single compression in cone channel) | 2.5 mL blood/min | Yes (ROS assay; slight effect on larger stiffer cells) | Yes (single cells deterministically printed into wells) | Yes (Raman identification demonstrated) |
| **Category 2: Continuous hydrodynamic displacement** | Real-time transit-time-based deformability-activated hydrodynamic sorting on a microfluidic chip | Microfluidic deformability-activated sorting of single particles.[24] | Direct contact and squeezing through a microconstriction for deformability sensing, followed by active hydrodynamic deflection at the sorting junction | ~600 particles/min (≈100 ms per particle) | Not reported | Yes; particles are actively routed to collection outlets rather than only characterized | Potentially yes in principle because target particles are collected at designated outlets, but not demonstrated for biological downstream assays in this paper |
| | Deformability-based cell margination in a straight microchannel with a three-outlet collection system | Deformability based cell margination—A simple microfluidic design for malaria-infected erythrocyte separation.[25] | Indirect cell-cell interaction–driven lateral displacement in concentrated blood flow inside a 15 µm × 10 µm straight microchannel; no constriction trapping step | 5 mL/min, approximately 20 million cells/min | Not reported | Yes; margination-enriched cells were continuously removed through the two side outlets of the 3-outlet design | Not demonstrated in this study, although the paper states the outlet flow is compatible with downstream detection such as Giemsa staining |

| | | | | | | | | |
|---|---|---|---|---|---|---|---|---|
| | Diagonal-ridge microfluidic viscoelastic sorting through repeated compression and relaxation | Microfluidic cellular enrichment and separation through differences in viscoelastic deformation.[26] | Direct repeated compression through ridge-defined gaps, combined with lateral translation in a ridged patterned channel | Not reported | Not reported | Yes; separated cell fractions were collected from different outlets and analyzed by flow cytometry | Not demonstrated |
| | Diagonal-ridge microfluidic stiffness-dependent separation by repeated compression and transverse migration | Stiffness Dependent Separation of Cells in a Microfluidic Device.[27] | Direct repeated compression by periodic diagonal ridges in a patterned microchannel, combined with ridge-induced secondary flow | 250 cells/s at 0.05 mL/min and 10^6 cells/mL | Yes; outlet-collected untreated K562 cells showed no significant difference in viable cell numbers after 6 days of culture compared with control | Yes; separated cells were continuously collected at the two outlets | Yes; outlet fractions were analyzed by flow cytometry, AFM, and post-flow cell culture |
| | Circular-microchannel inertial-position analysis of suspended cells with cytochalasin D perturbation | The influence of cell elastic modulus on inertial positions in Poiseuille microflows.[28] | Noncontact advection in circular Poiseuille microflow; cells migrate to characteristic inertial positions across the channel diameter | Not reported | Not reported | Not demonstrated | Not demonstrated |
| | Ridge-based microfluidic stiffness sorting of chemotherapy-treated cells using 3-outlet and 5-outlet devices | Microfluidic cell sorting by stiffness to examine heterogenic responses of cancer cells to chemotherapy.[29] | Direct repeated compression and transverse migration in microchannels containing periodic diagonal ridges | ~500 cells/s at 0.03 mL/min for the 3-outlet sorting experiments | Yes, functionally demonstrated after sorting; for K562 treated with 2 µM daunorubicin, soft outlet viability = 94.7% and stiff outlet viability = 7.3% | Yes; cells were collected from soft, middle, and stiff outlets for further analysis | Yes; sorted cells were analyzed by AFM, flow cytometry, qPCR/PCR array, network analysis, and inhibitor validation |
| | Diagonal-ridge microfluidic channel for continuous stiffness-based separation | Designing microfluidic channel that separates elastic particles upon stiffness.[30] | Direct mechanical interaction with rigid diagonal ridges during passage through narrow constrictions; importantly, no adhesive interaction is required | Not reported; proposed as a continuous-flow separation design based on simulation, but no experimental throughput value is given | Not reported | Not demonstrated | Not demonstrated |
| | Deformability activated cell sorting (DACS) using inertial microfluidics | Deformability-based cell classification and enrichment using inertial microfluidics.[31] | Predominantly hydrodynamic, contactless lateral focusing/migration in a straight high-aspect-ratio channel, followed by collection through branched outlets | Single device throughput ~22,000 cells/min when operated at Rc = 42 | Processed MCF7 cells remained highly viable, similar to controls; viability plots on Fig. 5 show about 91–93% same day and 75–78% after 24 h | Yes; enriched fractions were collected off-chip from designated cancer outlets, with 96% recovery for modMCF7 and 97% yield for SAOS-2 at the collection outlets under optimal conditions | Yes; authors performed global gene expression (microarray) on processed cells and concluded there were no significant gene-expression alterations at 2-fold change; they also state the approach is suitable for gene expression analysis and establishment of in vitro culture |
| | Pinched flow fractionation (PFF) for cancer cell/WBC separation | Separation of cancer cells from white blood cells by pinched flow fractionation. [32] | Cells are hydrodynamically pinched against the side wall in a narrow segment, where wall-induced deformation effects influence separation. | Optimal separation occurred at 10 µL $h^{-1}$; higher flow rates decreased WBC removal. | Not directly measured/reported; authors state viability is not expected to change because exposure to high shear is very short | Yes; separated cells were collected in small-particle outlet and large-particle outlets, with drain mainly collecting buffer/aggregates | Potentially yes, because the paper is framed as CTC enrichment that should enable further analysis, but no downstream molecular/functional assay was directly demonstrated in this study |
| **Category 3: Noncontact or low-contact** | Extensional flow device for apoptotic bodies | Non-contact microfluidic mechanical property measurements of single apoptotic bodies. [33] | Noncontact (stretched in extensional flow) | 0.5–2 cells/s (≈ 1 000–5 000 per assay) | ≈ 100 % | Yes | Yes |
| | Suspended microchannel resonator (SMR) | Deformability of tumor cells versus blood cells. [34] | Noncontact (no squeezing) | 0.01–0.1 cells/s (30–300 cells/h) | ≈ 100 % | Yes | Yes |
| | Contactless hydro-stretching DC (lh-DC) and contact constriction DC (cc-DC) | A Systematic Study of Size Correlation and Young's Modulus Sensitivity for Cellular Mechanical Phenotyping by Microfluidic Approaches.[35] | lh-DC: Noncontact; cc-DC: Direct contact through constriction | lh-DC: 140 cells/s; cc-DC: 180 cells/s | Not quantitatively reported | Yes (cells remain intact after measurement) | No |
| | Suspended microchannel resonator (SMR) with integrated constriction for deformability and surface friction analysis | Characterizing deformability and surface friction of cancer cells.[36] | Direct contact and squeezing through a 6 µm-wide, 50 µm-long constriction; cell-wall interaction is integral to measurement | A few thousand cells/h for single-cell buoyant mass + passage analysis; up to ~10^5 cells/h in blood-spiking discrimination mode | Yes; cells remained viable and proliferated well after SMR measurement | Potentially yes; cells remain viable after assay, though this paper focuses on characterization rather than active physical sorting/recovery workflow | Yes |
| | Cross-flow extensional microfluidic deformation assay in shear- and inertia-dominant regimes | Cells Under Stress: An Inertial-Shear Microfluidic Determination of Cell Behavior.[37] | Noncontact hydrodynamic deformation at the stagnation point of an extensional flow | Not reported | Yes, condition-dependent: viability remained within error of undeformed control for $Q \leq 600$ µL/min in the inertial regime, but dropped to <50% for $Q > 600$ µL/min | Not demonstrated | Not demonstrated in this study, although the paper states cells below the failure point could continue to be studied |
| | Cross-slot microfluidic extensional-flow assay with analytical extraction of cell viscoelastic parameters | Measuring Cell Viscoelastic Properties Using a Microfluidic Extensional Flow Device.[38] | Noncontact hydrodynamic stretching near the stagnation point of planar extensional flow | Not reported | Not reported | Not demonstrated | Not demonstrated |

| | | | | | | | | |
|---|---|---|---|---|---|---|---|---|
| | Computationally corrected cross-slot microfluidic deformability cytometry using elongation index (EI) | Elongation Index as a Sensitive Measure of Cell Deformation in High-Throughput Microfluidic Systems.[39] | Noncontact hydrodynamic stretching in extensional flow at the stagnation point of a cross-slot microchannel | No new experimental throughput reported; the paper describes mDC as operating at up to 2000 cells/s | Not reported | Not demonstrated | Not demonstrated |
| | Inertial microfluidic cell stretcher (iMCS) using inertial focusing and T-junction wall-impact stretching with near real-time image analysis | Inertial Microfluidic Cell Stretcher (iMCS): Fully Automated, High-Throughput, and Near Real-Time Cell Mechanotyping. [40] | High-momentum deformation upon collision with a rigid PDMS wall at a T-junction after inertial focusing and spacing control | ~450 cells/s actual rate (~1000 cells/s theoretical), with up to 6300 cells analyzed in 3 s and a total run time of ~35 s. | Yes, the paper states comparable cell viability after cell-wall collision and no cell-wall adhesion even after an average of >18 000 impact events | Yes, the Discussion states processed cells can be collected | Potentially yes / discussed, but not shown as a dedicated post-recovery assay workflow in the main experiments |

**Table S2. Correlation analysis between cell stiffness and structural protein expression across the tested cell lines.** Spearman correlation analysis between cell stiffness measured by Stiff-FCS and the expression intensities of Vimentin and Lamin A/C in the tested cell lines. Correlation coefficients (ρ) and corresponding P values are listed for each cell line to evaluate the association between structural protein expression and cell stiffness at the single-cell level.

| | *Vimentin* | | *Lamin A/C* | | *AR* | | *Solidity* | | *Cell Number* |
|---|---|---|---|---|---|---|---|---|---|
| | Spearman | p value | Spearman | p value | Spearman | p value | Spearman | p value | |
| *MDA-MB-231* | 0.173 | 0.000 | 0.272 | 0.000 | -0.241 | 0.000 | 0.316 | 0.000 | 1029 |
| *MCF7* | -0.005 | 0.921 | 0.337 | 0.000 | 0.453 | 0.000 | -0.116 | 0.029 | 354 |
| *HCC1806* | 0.116 | 0.095 | 0.382 | 0.000 | 0.398 | 0.000 | -0.021 | 0.768 | 208 |
| *HCC70* | 0.142 | 0.039 | 0.206 | 0.003 | 0.086 | 0.213 | -0.035 | 0.610 | 212 |
| *H1299* | 0.188 | 0.000 | 0.140 | 0.002 | -0.146 | 0.001 | 0.183 | 0.000 | 487 |
| *H3122* | -0.255 | 0.000 | 0.280 | 0.000 | -0.063 | 0.021 | 0.025 | 0.370 | 1336 |
| *PY8119* | 0.022 | 0.437 | 0.031 | 0.263 | -0.266 | 0.000 | 0.172 | 0.000 | 1297 |
| *Fibroblast* | 0.215 | 0.000 | -0.006 | 0.878 | 0.066 | 0.118 | 0.134 | 0.001 | 568 |

**Table S3. Summary statistics of single-cell stiffness across four HNC specimens measured by Stiff-FCS.** Summary statistics of single-cell stiffness measured by Stiff-FCS for each HNC specimen. For each cell line, the total number of analyzed cells (n), mean stiffness, standard deviation (SD), median stiffness, first quartile (Q1), third quartile (Q3), and interquartile range (IQR) are reported.

| *Specimen* | *Cell number (n)* | *Mean stiffness (kPa)* | *SD (kPa)* | *Median stiffness (kPa)* | *Q1 (kPa)* | *Q3 (kPa)* | *IQR (kPa)* |
|---|---|---|---|---|---|---|---|
| *JHU029* | 492 | 0.56 | 1.62 | 0.40 | 0.27 | 0.57 | 0.30 |
| *SCC47* | 743 | 2.11 | 3.49 | 1.23 | 0.83 | 2.12 | 1.29 |
| *HN6* | 755 | 0.83 | 1.35 | 0.59 | 0.34 | 0.92 | 0.58 |
| *HN12* | 473 | 1.97 | 2.80 | 1.04 | 0.33 | 2.62 | 2.29 |